\documentclass[aps,prd,nofootinbib,showpacs,superscriptaddress,floatfix, preprint]{revtex4-1}
\pdfoutput=1
\usepackage[utf8]{inputenc}
\usepackage{amsmath,amssymb}
\usepackage{epsfig}
\usepackage{graphicx}
\usepackage[usenames,dvipsnames]{color}
\usepackage[colorlinks,citecolor=blue]{hyperref}
\usepackage{color}
\usepackage{subfigure}
\usepackage{hhline}
\usepackage{multirow}

\begin{document}
	\title{Revisiting Majorana Neutrino Textures in the Light of Dark LMA}
	
	\author{Happy Borgohain}
	\email{happy@iitg.ac.in}

	\author{Debasish Borah}
	\email{dborah@iitg.ac.in}
	\affiliation{Department of Physics, Indian Institute of Technology Guwahati, Assam 781039, India}

	\begin{abstract}
	We study the possibility of texture zeros in Majorana light neutrino mass matrix in the light of dark large mixing angle (DLMA) solution to solar neutrino problem where solar mixing angle ($\sin^2{\theta_{12}}\simeq 0.7 $) lies in the second octant instead of first octant in standard large mixing angle (LMA) scenario ($\sin^2{\theta_{12}}\simeq 0.3 $). In three neutrino scenario, we find that LMA and DLMA solutions lead to different set of allowed and disallowed textures with one and two zeros. While being consistent with existing bounds from neutrino oscillation data, neutrinoless double beta decay and cosmology these allowed textures also lead to interesting correlations among light neutrino parameters which can distinguish LMA from DLMA solution. We also check the implications for texture zeros in $3+1$ neutrino scenario using both LMA and DLMA solutions. While LMA and DLMA solutions do not play decisive role in ruling out texture zeros in this case, they do give rise to distinct predictions and correlations between light neutrino parameters.
	
		\end{abstract}
	\maketitle

\section{Introduction}
	
The fact that neutrinos have tiny but non-zero mass and large mixing have been established due to irrefutable amount of evidences gathered in last few decades \cite{Mohapatra:2005wg, Tanabashi:2018oca}. Three non-zero mixing angles and two mass squared differences have been measured upto unprecedented accuracy in recent neutrino oscillation experiments upto a certain caveats. For example, the octant of atmospheric mixing angle, mass ordering, Dirac CP phase are not yet settled. In addition to these, the nature of light neutrinos: Majorana or Dirac, lightest neutrino mass also remain undetermined at neutrino oscillation experiments. For a recent global fit of three neutrino oscillation data, we refer to \cite{deSalas:2017kay, Esteban:2018azc}. If neutrinos are Majorana fermions, two more phases appear in three neutrino scenarios which can be probed only at alternative experiments like neutrinoless double beta decay $(0\nu \beta \beta)$. Apart from oscillation experiments, neutrino sector also gets constrained from cosmology due to the upper bound on sum of absolute neutrino masses from Planck 2018 data $\sum_i \lvert m_i \rvert < 0.12$ eV \cite{Aghanim:2018eyx}.


The above mentioned experimental input on light neutrino parameters are based on the assumption that light neutrinos interact with matter purely via standard model (SM) interactions. However, if in addition to the standard interactions, neutrinos have subdominant non standard interactions (NSI) with the matter fields, determining the neutrino parameters will go through new ambiguities. The idea of NSI was first introduced by Wolfenstein in 1978 in his landmark paper that also identified the
conventional matter effect \cite{Wolfenstein:1977ue}  and then subsequently in \cite{Guzzo:1991hi}, \cite{Roulet:1991sm} to account for the possible alternative solution to the solar neutrino problem. Since then, substantial efforts have been put to study its possible consequences. Like the standard interactions, NSI can also be divided into two groups, neutral current NSI (NCNSI) and charged current NSI (CCNSI). NCNSI and CCNSI affects the neutrino propagation in matter and the production and detection of neutrinos respectively. Both CCNSI and NCNSI are extensively studied in literature providing some lower limit on the value of the couplings in order to have  a resolvable impact on the upcoming oscillation experiments. The presence of NSI in neutrino propagation may give rise, among other effects, to a degeneracy in the measurement of the solar mixing angle, $\theta_{12}$. Although the large mixing angle (LMA) solution ($\Delta m_{21}^2 \simeq 7.5 \times 10^{-5}, \sin^2{\theta_{12}}\simeq 0.3 $) is mostly considered as the solution to the solar neutrino problems, the presence of NSI there exists a nearly degenerate solution for other octant of solar mixing angle ($\theta > \pi/4$), i.e., in the second octant ($\Delta m_{21}^2 \simeq 7.5 \times 10^{-5}, \sin^2{\theta_{12}}\simeq 0.7 $). This degenerate solution to the solar mixing problem is often referred to as dark LMA (DLMA) solution in the literature \cite{Miranda:2004nb, Escrihuela:2009up, Farzan:2017xzy}. Several studies have been done in the context of this DLMA solution, for example, the impact of DLMA in determining neutrino mass hierarchy at neutrino oscillation experiments \cite{Farzan:2017xzy, Bakhti:2014pva, Coloma:2016gei, Deepthi:2016erc, Choubey:2019osj}, the impact of DLMA on $0\nu \beta \beta$ lifetime with three neutrinos \cite{N.:2019cot}, $3+1$ neutrinos \cite{Deepthi:2019ljo}. Further studies related to resolving the degeneracy were done by the authors of \cite{Escrihuela:2009up, Coloma:2017egw} while the constraints from COHERENT experiment (coherent neutrino nucleus scattering data) on DLMA solution were studied in \cite{Coloma:2017ncl, Denton:2018xmq}. In spite of stringent constraints on neutrino NSI, the recent global fit including oscillation and COHERENT data \cite{Esteban:2018ppq} still allows DLMA solution at $3\sigma$ level.

Motivated by the recent interest in DLMA solution, here we study its implications for texture zeros in neutrino mass matrix. If neutrinos are of Majorana type, as we assume in our work, the $3\times 3$ mass matrix has six independent complex parameters equivalent to twelve real parameters. On the other hand, we have experimental input on five parameters only namely, three mixing angle and two mass squared differences. Similar situation arises in neutrino mass models too where there are many free parameters. However, in the presence of some underlying symmetries, the mass matrix can have very specific structure reducing the number of free parameters thereby enhancing its predictive power. In such a case, we can have very specific predictions for light neutrino parameters like CP phase, octant of atmospheric mixing angle, mass ordering which can be tested at ongoing experiments. Here we consider such a possibility where an underlying symmetry can restrict the mass matrix to have non-zero entries only at certain specific locations. Known as texture zero models in the literature, a review of such scenarios within three neutrino framework can be found in \cite{Ludl:2014axa} \footnote{Also see \cite{Xing:2002ta, Merle:2006du,  Singh:2016qcf, Ahuja:2017nrf, Borah:2015vra, Kalita:2015tda, Borgohain:2018lro, Borgohain:2019pya} for texture related works in different contexts.}. Without considering any UV completion based on symmetries that give rise to such textures, we focus on their phenomenology specially with respect to comparison between LMA and DLMA from texture zero predictions. In the diagonal charged lepton basis, if the light neutrino mass matrix has some zeros, one gets the corresponding number of constraints relating light neutrino parameters. Solving the texture zero equations lead to predictions of light neutrino parameters. Such predictions for known parameters must satisfy experimental bounds while the same for unknown parameters can be tested at upcoming or complementary experiments. Such predictions can be used to discriminate between different textures as have been already worked out in several earlier works. Here we not only compare different texture zero mass matrices but also compare the consequences of standard LMA and DLMA solutions for texture zero models. It has already been shown in earlier works that in the diagonal charged lepton basis, not more than two zeros are allowed in the light neutrino mass matrix. While all six possible one zero texture $(^6C_n, n=1)$ are allowed, among the fifteen possible two zero textures, only six were found to be allowed after incorporating both neutrinos as well as cosmology data \cite{Meloni:2014yea,Fritzsch:2011qv,Alcaide:2018vni,Zhou:2015qua,Bora:2016ygl,Borgohain:2018lro}. We first make a list of allowed and disallowed one zero and two zero texture mass matrices for LMA as well as DLMA scenarios and compare the predictions for light neutrino parameters. We also check the viability from cosmological bound on sum of absolute neutrino masses as well as experimental lower bounds on $0\nu \beta \beta$ lifetime. In the end, we also study the consequence of DLMA solution for Majorana neutrino textures by considering  $3+1$ neutrino framework in view of short baseline neutrino anomalies from LSND \cite{Aguilar:2001ty} and MiniBooNE \cite{AguilarArevalo:2007it,AguilarArevalo:2010wv,Aguilar-Arevalo:2018gpe} experiments suggesting the presence of additional light neutrino at eV scale.

The paper is organised as follows. In section \ref{sec:level2} we discuss the texture zero mass matrices in three neutrino scenario. We briefly discuss neutrinoless double beta decay in section \ref{sec:level3} . In section \ref{sec:level5} we discuss our results of three neutrino scenario in details. We discuss texture zeros in $3+1$ neutrino case in section \ref{sec:level6} and finally conclude in section \ref{sec:level7}.

\section{Texture Zero Mass Matrices}{\label{sec:level2}}
As mentioned earlier, texture zeros in lepton mass matrices increase the predictive power of the model due to a decrease in the number of free parameters \cite{Ludl:2014axa,Xing:2002ta,Frampton:2002yf, Merle:2006du,Singh:2016qcf,Ahuja:2017nrf,Meloni:2014yea,Fritzsch:2011qv,Alcaide:2018vni,Zhou:2015qua,Bora:2016ygl,Borgohain:2018lro}. The zero texture models are widely studied as the number of free parameters can be significantly reduced in such models . It has been shown that in the diagonal charged lepton basis, not more than two zeros are allowed in the light neutrino mass matrix. There are $^6C_1=6$ and $^6C_2=15$ classes of possible  one-zero and two-zero texture neutrino mass matrices. In previous studies, out of 15 possible two-zero texture neutrino mass matrices, only 7 were shown to be allowed by experimental datas which are being named as A1, A2, B1, B2, B3, B4, C1 below. However, previous studies were based on LMA solution only. Therefore, here we check the validity of all 15 two-zero textures using both LMA as well as DLMA solution. On the other hand, due to the less restrictive nature, all six one-zero texture mass matrices were found to be allowed in previous studies. The one-zero texture neutrino mass matrices are named as G1, G2, G3, G4, G5 and G6, The two-zero (equations \eqref{eq1}-\eqref{eq3c}) and one-zero (equations \eqref{eq4}-\eqref{eq5}) neutrino mass matrices can be written as,

\begin{equation}\label{eq1}
A1=\left(\begin{array}{ccc}
0 & 0 & \times\\
0& \times& \times\\
\times & \times& \times
\end{array}\right),A2=\left(\begin{array}{ccc}
0 & \times &0 \\
\times & \times& \times\\
0& \times& \times
\end{array}\right)
\end{equation}

\begin{equation}\label{eq2}
B1=\left(\begin{array}{ccc}
\times & \times &0\\
\times& 0& \times\\
0 & \times& \times
\end{array}\right),B2=\left(\begin{array}{ccc}
\times & 0&\times \\
0 & \times& \times\\
\times& \times& 0
\end{array}\right), B3=\left(\begin{array}{ccc}
\times & 0& \times \\
0 & 0 & \times\\
\times & \times & \times
\end{array}\right), B4=\left(\begin{array}{ccc}
\times & \times& 0 \\
\times & \times & \times\\
0 & \times & 0
\end{array}\right)
\end{equation}

\begin{equation}\label{eq3}
C1=\left(\begin{array}{ccc}
\times & \times & \times\\
\times& 0& \times\\
\times & \times& 0
\end{array}\right)
\end{equation}

\begin{equation}\label{eq3a}
D1=\left(\begin{array}{ccc}
\times & \times &\times\\
\times& 0& 0\\
\times & 0& \times
\end{array}\right),D2=\left(\begin{array}{ccc}
\times & \times&\times \\
\times & \times& 0\\
\times& 0& 0
\end{array}\right)
\end{equation}

\begin{equation}\label{eq3b}
E1=\left(\begin{array}{ccc}
0 & \times &\times\\
\times& 0& \times\\
\times & \times& \times
\end{array}\right),E2=\left(\begin{array}{ccc}
0 & \times&\times \\
\times & \times& \times\\
\times& \times& 0
\end{array}\right), E3=\left(\begin{array}{ccc}
0 & \times&\times \\
\times & \times & 0\\
\times & 0 & \times
\end{array}\right)
\end{equation}

\begin{equation}\label{eq3c}
F1=\left(\begin{array}{ccc}
\times & 0 &0\\
0& \times& \times\\
0 & \times& \times
\end{array}\right),F2=\left(\begin{array}{ccc}
\times & 0&\times \\
0 & \times& 0\\
\times& 0& \times
\end{array}\right), F3=\left(\begin{array}{ccc}
\times & \times& 0 \\
\times & \times & 0\\
0 & 0 & \times
\end{array}\right)
\end{equation}

\begin{equation}\label{eq4}
G1=\left(\begin{array}{ccc}
0 & \times &\times\\
\times& \times& \times\\
\times & \times& \times
\end{array}\right),G2=\left(\begin{array}{ccc}
\times & 0&\times \\
0 & \times& \times\\
\times& \times& \times
\end{array}\right), G3=\left(\begin{array}{ccc}
\times & \times& 0\\
\times& \times & \times\\
0 & \times & \times
\end{array}\right), G4=\left(\begin{array}{ccc}
\times & \times& \times \\
\times & 0 & \times\\
\times & \times & \times
\end{array}\right)
\end{equation}

\begin{equation}\label{eq5}
G5=\left(\begin{array}{ccc}
\times & \times & \times\\
\times& \times& 0\\
\times & 0& \times
\end{array}\right),G6=\left(\begin{array}{ccc}
\times& \times &\times \\
\times & \times& \times\\
\times& \times& 0
\end{array}\right)
\end{equation}

where the crosses "$\times$'' denote non-zero arbitrary elements of light neutrino mass matrix. 

\section{Neutrinoless double beta decay}
\label{sec:level3}
As mentioned earlier, neutrinoless double beta decay is a process, if observed, can prove the Majorana nature of light neutrinos. It is a process where a nucleus emits two electrons thereby changing its atomic number by two units
	$$ (A, Z) \rightarrow (A, Z+2) + 2e^- $$
	with no neutrinos in the final state. Such a process violates lepton number by two units and hence is a probe of Majorana neutrinos, which are predicted by generic seesaw models of neutrino masses. For a review of $0\nu \beta \beta $ process and current limits, one may refer to \cite{Rodejohann:2011mu, Cardani:2018lje, Dolinski:2019nrj}. Apart from probing the Majorana nature of light neutrinos, observation of such a process can also discriminate between neutrino mass ordering: normal ordering (NO) vs inverted ordering (IO), different values of Majorana CP phases. With precise information on phase space factors (PSF) and associated nuclear matrix element (NME), it is possible to set tight constraints on the absolute neutrino mass scale using the lower bounds on $0 \nu \beta \beta$ half-life given by experiments like KamLAND-Zen \cite{KamLAND-Zen:2016pfg}. Among the recent experiments, this one quotes the most stringent lower bound on the half-life of $0 \nu \beta \beta$ using $^{136} Xe$ nucleus as $ \rm T_{1/2}^{0\nu}>1.07\times 10^{26}$ year at $ 90\%$ C. L. This can be translated to an upper limit of effective Majorana mass $\lvert m_{ee} \rvert$ in the range $ (0.061-0.165)$ eV where the uncertainty arises due to the NME. Although the net contribution to this process is model dependent, we stick to the minimal scenario where only the light neutrinos contribute to it. This standard contribution is mediated by purely left handed (LH) currents and the corresponding amplitude of the process is 
\begin{equation}\label{eqa1}
		\rm{A_\nu}^{LL} \propto G_{F}^2\sum_i \frac{{U^2_{{e_i}}}m_i}{p^2}=G_{F}^2 \frac{m_{ee}}{p^2}
		\end{equation}
		where, $\left|p\right|\sim$ 100 MeV is the typical momentum transfer at the leptonic vertex, U represents the leptonic mixing matrix, $m_i$ are the masses for the three generations of light Majorana neutrinos. The corresponding half-life is 
\begin{equation}\label{eq48}
	\left[{T_{\frac{1}{2}}}^{0\nu}\right]^{-1}=G^{0\nu}(Q,Z)\left({\left| M^{0\nu}_\nu\eta_\nu \right|}^2 \right),
	\end{equation}
	where $\eta_\nu$ contains the particle physics input to the process given by 
	\begin{equation}\label{eq43}
	\left|\eta_{\nu}\right|=\frac{1}{m_e}\sum_i U^2_{e_i} m_i
	\end{equation}
	In the above expression for half-life, $G^{0\nu}(Q,Z)$ represents the phase space factor and $M^{0\nu}$ is the nuclear matrix element mentioned earlier. The numerical values of these quantities for specific nuclei are shown in tabular form in table \ref{3} \cite{Dev:2014xea}.

	\begin{table}[h!]
		\centering
		\begin{tabular}{|c|| c|c| c| c| }
			\hline
			Isotope	&	$G^{0\nu}(Q,Z)(yr^{-1})$&$M^{0\nu}_\nu$  \\ \hline
			$^{76}Ge$	&	5.77$\times 10^{-15}$&2.58-6.64  \\ \hline\hline\hline
			$^{136}Xe$	&	3.56$\times 10^{-14}$&1.57-3.85  \\ \hline\hline\hline		
		\end{tabular}
		\caption{The different values of PSF and NME for different nuclei used in NDBD experiments.} \label{3}
	\end{table}

	\section{Results and Discussion}{\label{sec:level5}}
	
	We first check the validity of different texture zero mass matrices from neutrino oscillation data. To solve the constraint equations corresponding to the texture zero conditions, we first parametrise the neutrino mass matrix in the 3$\nu$ scheme as,
	\begin{equation}\label{eq6}
	M_\nu= U_{\rm PMNS}{M_\nu}^{(\rm diag)} {U_{\rm PMNS}}^T,
	\end{equation}
	where, $U_{\rm PMNS}=U$ is the usual Pontecorvo-Maki-Nakagawa-Sakata (PMNS) mixing matrix. In general, the PMNS mixing matrix consists of the diagonalising matrix of the neutrino and charged lepton mass matrices as,
	\begin{equation}\label{eq7}
	U_{\text{PMNS}} = U^{\dagger}_l U_{\nu}
	\end{equation} 
	In the diagonal charged lepton basis $U_{\rm PMNS}=U_{\nu}$. The PMNS mixing matrix can be parametrised in terms of the leptonic mixing angles and phases as
	\begin{equation}\label{eq8}
	U=U_{\text{PMNS}}=\left[\begin{array}{ccc}
	c_{12}c_{13}&s_{12}c_{13}&s_{13}e^{-i\delta}\\
	-c_{23}s_{12}-s_{23}s_{13}c_{12}e^{i\delta}& c_{23}c_{12}-s_{23}s_{13}s_{12}e^{i\delta}&s_{23}c_{13}\\
	s_{23}s_{12}-c_{23}s_{13}c_{12}e^{i\delta}&-s_{23}c_{12}-c_{23}s_{13}s_{12}e^{i\delta}&c_{23}c_{13}
	\end{array}\right]P
	\end{equation}
	where $c_{ij} = \cos{\theta_{ij}}, \; s_{ij} = \sin{\theta_{ij}}$ and $\delta$ is the leptonic Dirac CP phase. The diagonal matrix $P=\text{diag}(1, e^{i\alpha}, e^{i(\beta+\delta)})$  contains the Majorana CP phases $\alpha, \beta$ that appears when $\nu$ is Majorana and are not constrained by neutrino oscillation data but has to be probed by alternative experiments. Also, in the above expression for $M_{\nu}$, the diagonal light neutrino mass matrix is denoted by ${M_\nu}^{(\rm diag)}= \textrm{diag}(m_1,m_2,m_3)$ where the light neutrino masses can follow either normal ordering (NO) or inverted ordering (IO). For NO, the three neutrino mass eigenvalues can be written as 
$$M^{\text{diag}}_{\nu}
= \text{diag}(m_1, \sqrt{m^2_1+\Delta m_{21}^2}, \sqrt{m_1^2+\Delta m_{31}^2})$$ while for IO, they can be written as 
$$M^{\text{diag}}_{\nu} = \text{diag}(\sqrt{m_3^2+\Delta m_{23}^2-\Delta m_{21}^2}, \sqrt{m_3^2+\Delta m_{23}^2}, m_3)$$ 
The analytical expressions of the elements of this mass matrix are given in Appendix \ref{appen1}. 

	\begin{table}[h!]
		\centering
		\begin{tabular}{||c|  c||}
			\hline
			PARAMETERS & $3 \sigma$ RANGES  (NO/IO)\\ \hline\hline\hline
			$\Delta m^2_{21} [10^{-5} \rm eV^2]$ & 6.79-8.01/6.79-8.01\\ \hline\hline
			$ \lvert \Delta m^2_{3l}\rvert[10^{-3}  \rm eV^2]$ & 2.432-2.618/2.416-2.603\\\hline\hline
			$\sin^2{\theta_{12}} (\rm LMA)$ & 0.275-0.350/0.275-0.350 \\ \hline\hline
			$\sin^2{\theta_{23}}$ & 0.427-0.609/0.430-0.612 	\\ \hline\hline
		    $\sin^2{\theta_{13}}$ & 0.02046-0.02440 /0.02066-0.02461\\ \hline\hline
		    
		\end{tabular}
		\caption{Global fit 3$\sigma$ values of $\nu$ oscillation parameters \cite{Esteban:2018azc}. Here $\Delta m^2_{3l} \equiv \Delta m^2_{31} {\rm (NO)}, \Delta m^2_{3l} \equiv \Delta m^2_{32} {\rm (IO)}$.}  \label{t0}
	\end{table}

From the parametrisation of the light neutrino mass matrix, it is clear that the $3\times 3$ Majorana neutrino mass matrix has nine independent parameters: three masses, three mixing angles and three phases. Out of these nine parameters, only five parameters namely, two mass squared differences and three mixing angles are measured at neutrino oscillation experiments, upto some ambiguity in determining the octant of $\theta_{23}$ mentioned earlier. For the one-zero texture mass matrices, we solve the two real equations corresponding to the texture zero condition and determine the parameter space in terms neutrino parameters. While solving these equations, we vary the lightest neutrino mass in the range $10^{-5}-0.1$ eV and the Dirac CP phase in the range $-\pi < \delta <\pi$.  For two-zero texture neutrino mass matrix, we have four real equations equating two independent complex elements to zero. Thus we can determine four unknown parameters out of the nine independent parameters of the neutrino mass matrix. Varying all the known neutrino oscillation parameters in their 3$\sigma$ range, we solved for the Majorana phases $\alpha$ and $\beta$, the Dirac CP phase $\delta$ and the lightest neutrino mass $m_1 (m_3)$ for NO (IO). For the solar mixing angle, we considered the standard LMA and the DLMA solutions and check the differences in resulting solutions of texture zero equations. It was extensively shown in \cite{Esteban:2018ppq} that the recent neutrino oscillation data \cite{Esteban:2018azc} and COHERENT data perfectly allows the DLMA solution at the 3$\sigma$ level for a smaller range of the NSI parameters and light mediator mass responsible for NSI heavier than about 10 MeV . In the presence of NSI, there is only a minute change of the parameters ${\sin}^2\theta _{12}$ and $\Delta m^2_{21}$ while the range of the other neutrino parameters for the 3 $\nu$ scenario are still stable. Thus we have used the global fit data as given in table \ref{t0} for our analysis. While for  ${\sin}^2\theta _{12}$, the range of values we have used for LMA and DLMA solutions in the 3 $\nu$ scenario (as in \cite{N.:2019cot}) are (0.275-0.350) and (0.650-0.725) respectively. We first check the validity of all possible one-zero and two-zero textures for both LMA and DLMA scenario and list the allowed and disallowed cases in table \ref{t1}. Here we implement only the neutrino oscillation data as constraints. Later we will implement the bounds from cosmology as well as neutrinoless double beta decay (NDBD). We implement these constraints one at a time in order to show the constraint which rules out a particular texture. As can be seen from table \ref{t1}, nine two-zero textures are completely ruled out by neutrino oscillation data alone for both LMA as well as DLMA while the other six two-zero textures namely, A1, B1, B2, B3, B4, C1 are allowed. Out of these six, while A1 is allowed with LMA only for NO of neutrino masses, C1 is allowed only with IO of light neutrino masses. The remaining four allowed textures do not discriminate between mass ordering as well as LMA, DLMA. Thus, one allowed texture (A1) show discrimination between  LMA, DLMA and two allowed textures (A1, C1) show discrimination between mass ordering in two-zero texture scenario. On the other hand, out of six different one-zero textures, G1 is allowed with LMA only for NO which is expected as G1 one-zero texture is a subclass of A1 two-zero texture. Also, the fact that G1 is allowed only with LMA and NO of light neutrino masses out of four different possibilities agree with the results of \cite{N.:2019cot} where they showed that NDBD amplitude can be vanishing only for LMA with NO of light neutrino masses. Out of the one-zero textures, G2, G3, G4, G6 are allowed for both the mass orderings as well as LMA, DLMA. The remaining one-zero texture G5 is allowed only for IO of light neutrino mass with both LMA and DLMA.

\begin{figure}[h!]
	\centering
	\includegraphics[width=0.47\textwidth]{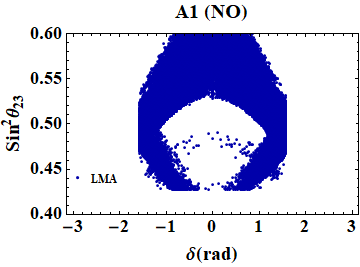}
	\includegraphics[width=0.47\textwidth]{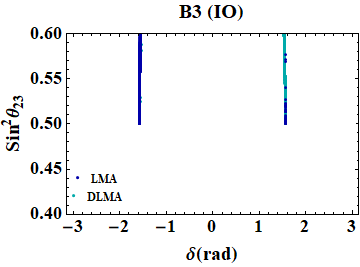}
	\includegraphics[width=0.47\textwidth]{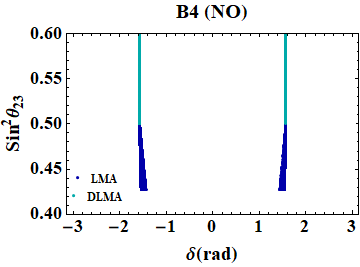}
	\includegraphics[width=0.47\textwidth]{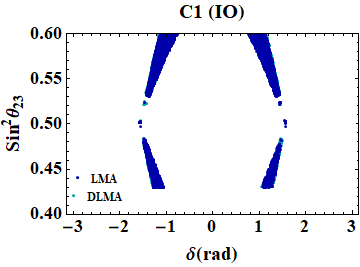}
	\includegraphics[width=0.47\textwidth]{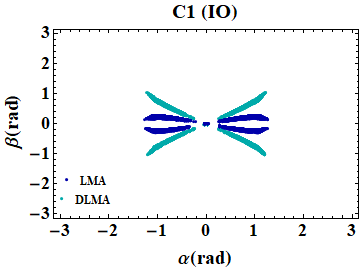}
	
	\caption{Correlations between light neutrino parameters for different allowed classes for two-zero texture.} \label{fig1}
\end{figure}

\begin{figure}[h]
	\includegraphics[width=0.47\textwidth]{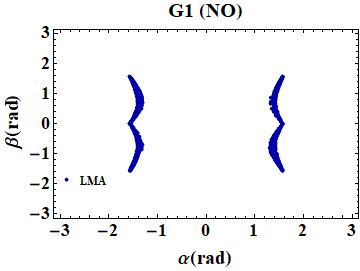}
	\includegraphics[width=0.47\textwidth]{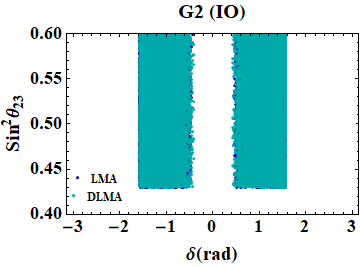}
		\includegraphics[width=0.47\textwidth]{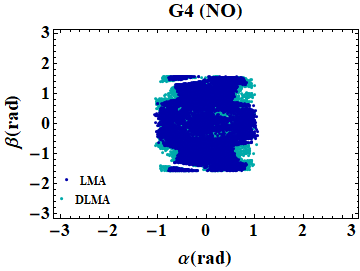}
	\caption{Correlations between light neutrino parameters for different allowed classes for one-zero texture.} \label{fig5}
\end{figure}

The analysis not only gives rise to a list of allowed and disallowed textures listed in table \ref{t1}, it also leads to some interesting correlations between light neutrino parameters dictated by the texture zero conditions. In particular, the predictions for unknown neutrino parameters like CP phases, octant of atmospheric mixing angle are of special importance. A few such correlations for two-zero and one-zero textures are shown in figure \ref{fig1} and figure \ref{fig5} respectively. While some of the textures predict a wide range of neutrino parameters, some of them predict very specific values of some parameters. For example, the two-zero textures B3, B4 predict maximal values of Dirac CP phase $\delta$. Similarly, B3 texture with IO prefers upper octant of atmospheric mixing angle. On the other hand B4 texture with NO shows different preference for atmospheric mixing angle with LMA and DLMA as seen from figure \ref{fig1}.
	
	\begin{table}[!htb]
		
		\begin{minipage}{.5\linewidth}
			
			\centering
			\begin{tabular}{||c||c |  c|| }
				\hline
				Class&	DLMA  & LMA  \\ \hline\hline\hline
			A1(NO/IO)&$ \times/ \times$	&$ \checkmark/ \times$  \\ \hline\hline
			A2(NO/IO)&$ \times/\times$	&$ \times/\times$ \\ \hline\hline
			B1(NO/IO)&$ \checkmark /\checkmark$ 	&$ \checkmark /\checkmark$\\ \hline\hline
			B2(NO/IO)&$ \checkmark /\checkmark$ 	&$ \checkmark/\checkmark$ \\ \hline\hline
			B3(NO/IO)&$ \checkmark /\checkmark$ 	&$ \checkmark /\checkmark$   \\ \hline\hline
			B4(NO/IO)&$ \checkmark /\checkmark$   &$\checkmark /\checkmark$\\ \hline	\hline
			C1(NO/IO)&$ \times /\checkmark$   &$\times/\checkmark$\\ \hline	\hline
			D1(NO/IO)&$ \times /\times$   &$\times /\times$\\ \hline	\hline
			D2(NO/IO)&$ \times /\times$   &$\times/\times$\\ \hline	\hline
			E1(NO/IO)&$ \times /\times$   &$\times /\times$\\ \hline	\hline
			E2(NO/IO)&$ \times /\times$   &$\times/\times$\\ \hline	\hline
			E3(NO/IO)&$ \times /\times$   &$\times /\times$\\ \hline	\hline
			F1(NO/IO)&$ \times /\times$   &$\times /\times$\\ \hline	\hline
			F2(NO/IO)&$ \times /\times$   &$\times /\times$\\ \hline	\hline
			F3(NO/IO)&$ \times /\times$   &$\times /\times$\\ \hline	\hline
				
			\end{tabular}
		\end{minipage}%
		\begin{minipage}{.5\linewidth}
			\centering
			
			\begin{tabular}{||c||c |  c|| }
				\hline
					Class&DLMA&	LMA    \\ \hline\hline\hline
				G1(NO/IO)&$ \times /\times$&$ \checkmark /\times$ \\ \hline\hline
				G2(NO/IO)&$ \checkmark/\checkmark$&$ \checkmark/\checkmark$	 \\ \hline\hline
				G3(NO/IO)&$ \checkmark /\checkmark$ &$ \checkmark /\checkmark$ \\ \hline\hline
				G4(NO/IO)&$ \checkmark /\checkmark$ &$ \checkmark /\checkmark$  \\ \hline\hline
				G5(NO/IO)&$ \times /\checkmark$ &$ \times /\checkmark$    \\ \hline\hline
				G6(NO/IO)&$ \checkmark /\checkmark$&$ \checkmark /\checkmark$  \\ \hline	\hline
				
			\end{tabular}
		\end{minipage} 
		\caption{Summary of allowed and disallowed two-zero textures (left) and one-zero textures (right) considering LMA and DLMA solutions. The $\checkmark$ or $\times$ symbol are used to denote if the class are allowed or disallowed by current experimental bounds.}\label{t1}
	\end{table}
	
\begin{figure}[h]
	\includegraphics[width=0.47\textwidth]{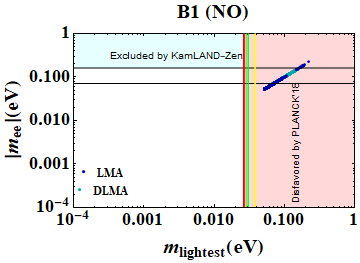}
	\includegraphics[width=0.47\textwidth]{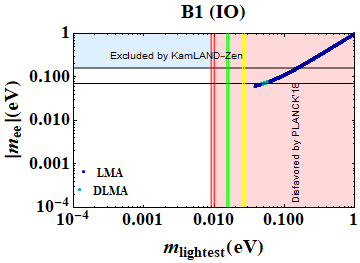}
	\includegraphics[width=0.47\textwidth]{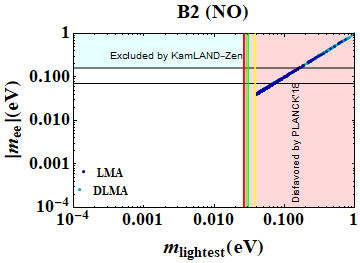}
	\includegraphics[width=0.47\textwidth]{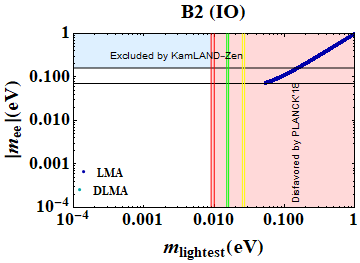}
		\caption{Effective Majorana neutrino mass governing NDBD as a function of the lightest neutrino mass for different allowed classes for two-zero textures. The three vertical lines  (red, green, yellow) corresponds to different sum of mass limits 0.11 eV, 0.12 eV, 0.14 eV respectively.} \label{fig4}
\end{figure}
\begin{figure}[h]
	\includegraphics[width=0.47\textwidth]{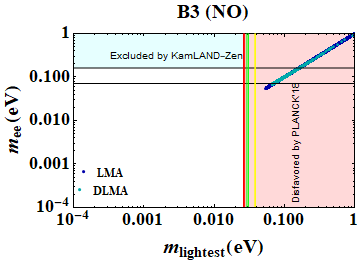}
	\includegraphics[width=0.47\textwidth]{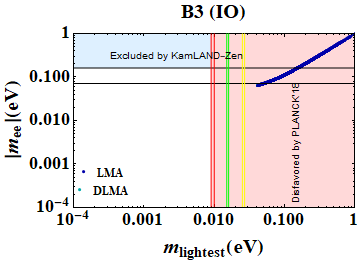}
	\includegraphics[width=0.47\textwidth]{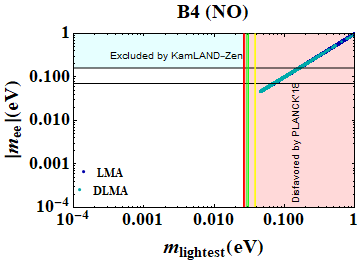}
	\includegraphics[width=0.47\textwidth]{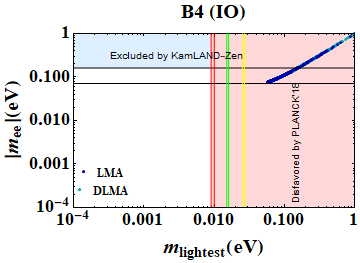}
	\includegraphics[width=0.47\textwidth]{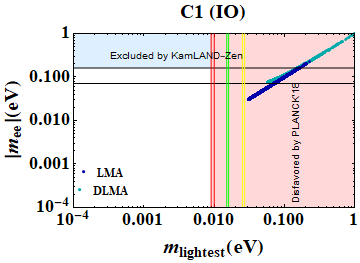}
	\caption{Effective Majorana neutrino mass governing NDBD as a function of the lightest neutrino mass for different allowed classes for two-zero textures. The three vertical lines  (red, green, yellow) corresponds to different sum of mass limits 0.11 eV, 0.12 eV, 0.14 eV respectively.} \label{fig4a}
\end{figure}
\begin{figure}[h]
	\includegraphics[width=0.47\textwidth]{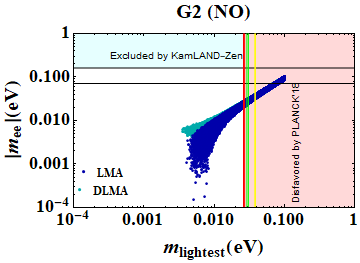}
	\includegraphics[width=0.47\textwidth]{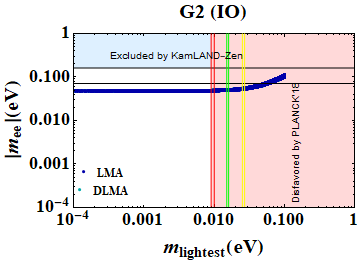}
	\includegraphics[width=0.47\textwidth]{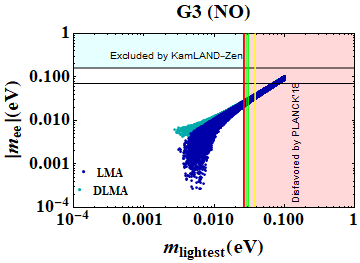}
	\includegraphics[width=0.47\textwidth]{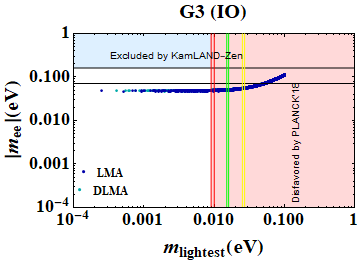}
		\caption{Effective Majorana neutrino mass governing NDBD as a function of the lightest neutrino mass for different allowed classes for one-zero textures.  The three vertical lines  (red, green, yellow) corresponds to different sum of mass limits 0.11 eV, 0.12 eV, 0.14 eV respectively.} \label{fig4b}
\end{figure}
\begin{figure}[h]	
	\includegraphics[width=0.47\textwidth]{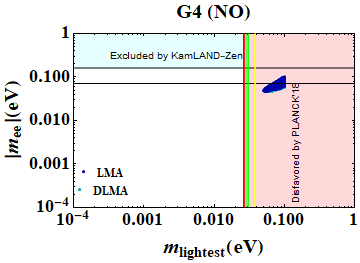}
	\includegraphics[width=0.47\textwidth]{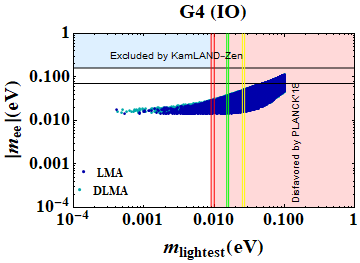}
	\includegraphics[width=0.47\textwidth]{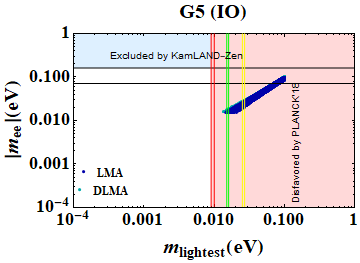}
	\includegraphics[width=0.47\textwidth]{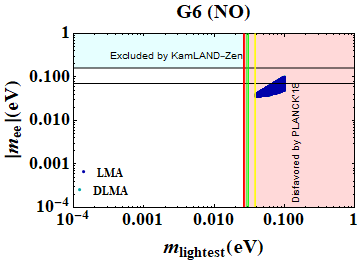}
	\includegraphics[width=0.47\textwidth]{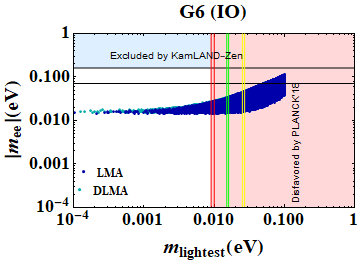}
	
	\caption{Effective Majorana neutrino mass governing NDBD as a function of the lightest neutrino mass for different allowed classes for one-zero textures. The three vertical lines  (red, green, yellow) corresponds to different sum of mass limits 0.11 eV, 0.12 eV, 0.14 eV respectively.} \label{fig4c}
\end{figure}

	\clearpage
	
	After checking the validity of texture zero mass matrices from neutrino oscillation data, we apply the constraints from neutrinoless double beta decay experiments. As discussed earlier, the neutrinoless double beta decay is governed by the term $m_{ee}$ known as the effective neutrino mass which can parameterised as
	\begin{equation}\label{eq9}
	m_{ee}=\sum_iU^2_{ei}m_i\hspace{1 mm},\hspace{1 mm}i=1,2,3,
	\end{equation}
	where, $U_{ei}, i=1,2,3$ is the first row of the PMNS mixing matrix given by equation \eqref{eq8}. In the standard parametrisation of the mixing matrix, $m_{ee}$ can be written as,
	\begin{equation}\label{eq10}
	\left|m_{ee}\right|= \lvert m_1{c^2_{12}}{c^2_{13}}+m_2{s^2_{12}}{c^2_{13}}e^{2i\alpha}+m_3{s^2_{13}}e^{2i\beta} \rvert.
	\end{equation}
Accordingly, the effective mass, as given by equation \eqref{eq10} can be expressed in terms of three unknowns in neutrino sector namely, the lightest neutrino mass $m_1 (m_3)$ and two Majorana phases $\alpha, \beta$. Figures \ref{fig4}, \ref{fig4a}, \ref{fig4b} and \ref{fig4c} show the effective mass governing NDBD as a function of the lightest neutrino mass for the two-zero and one-zero cases respectively which are allowed by neutrino oscillation data discussed earlier. We have considered the most stringent upper bound on the effective mass provided by the KamLAND-Zen experiment, i.e., $\lvert m_{ee} \rvert \leq (0.061-0.165)\; {\rm eV}$ \cite{KamLAND-Zen:2016pfg} shown as horizontal bands in figures \ref{fig4}, \ref{fig4a}, \ref{fig4b} and \ref{fig4c}. We also apply the cosmological upper bound on sum of absolute neutrino masses $\sum_i \lvert m_i \rvert < 0.11, 0.12, 0.14 \; {\rm eV}$  \cite{Aghanim:2018eyx} corresponding to the vertical lines of colour red, green and yellow respectively in the plots. The three bounds we have used corresponds to different datasets used in the analysis namely (Planck TT,TE,EE+lowE+lensing+BAO+Pantheon), (Planck TT,TE,EE+lowE+lensing+BAO) and (Planck TT,TE,EE+lowE+lensing+BAO+DES) data respectively all at 95$\%$ CL. We have translated the bound on sum of the absolute neutrino mass into the corresponding bound on the lightest neutrino mass, depicted by the rightmost region in the plots shown in figures \ref{fig4}, \ref{fig4a}, \ref{fig4b} and \ref{fig4c}. Each of these bounds on $\sum_i \lvert m_i \rvert$ correspond to two distinct exclusion lines in this plots. This is due to the $3\sigma$ values of mass squared differences used to find the corresponding lower bound on the lightest neutrino mass. Since the definition of the lightest neutrino mass is slightly different for NO and IO, we also get a little difference in the lower bound on $m_{\rm lightest}$ for NO and IO, as evident from the plots shown in figures \ref{fig4}, \ref{fig4a}, \ref{fig4b} and \ref{fig4c}. Clearly, almost all the two-zero textures allowed by neutrino oscillation data are now saturating the upper bound on effective neutrino mass from neutrinoless double beta decay experiments. While all of them are marginally allowed (at least for one of the mass orderings and LMA, DLMA scenarios) by NDBD constraints, they all are disfavoured by cosmological upper bound on lightest neutrino mass except class B2 which marginally satisfies the weaker version of cosmological upper bound bound $\sum_i \lvert m_i \rvert < 0.14$ eV for NO. The two-zero texture $A1$ gives rise to vanishing contribution to NDBD by definition while it remains still allowed from cosmology bound (with LMA). Among the one-zero textures, while most of them saturate the bounds from NDBD experiment for some part of parameter space, none of them gets completely ruled out by it. After applying the cosmological upper bound on the sum of absolute neutrino masses however, one of the one-zero textures get completely disfavoured as can be seen from figures \ref{fig4b} and \ref{fig4c}. Several of these textures also show interesting contrast between LMA and DLMA as far as contributions to NDBD amplitude is concerned. For example, among one-zero textures G2 (NO), G3 (NO) show interesting contrasts near $m_{\rm lightest} \sim 0.005$ eV. We have summarised the results after applying NDBD and cosmology bound on allowed two-zero and one-zero textures in table \ref{t2} and table \ref{t3}. Thus, out of two-zero textures only one of them A1 is allowed with NO and LMA. Among the one-zero textures only G2, G3 and G6 are allowed for both the hierarchies and LMA, DLMA while G4 and G5 are allowed only for IO but for both LMA, DLMA. On the other hand, G1 is allowed only with NO and LMA as mentioned earlier.

	 \begin{table}
	\centering
	\renewcommand{\arraystretch}{1.5}
	\begin{tabular}{|p{1.95cm}|p{1.6cm}|p{1.6cm}|p{1.6cm}|p{1.6cm}|p{1.6cm}|p{1.6cm}|p{1.6cm}|p{1.6cm}|}
		\hline
		\multirow{3}{1.5cm}{\textbf{Class}} &\textbf{NDBD (LMA)}&\textbf{NDBD (DLMA)}& \multicolumn{3}{c|}{\textbf{ COSMOLOGY (LMA)}} & \multicolumn{3}{c|}{\textbf{ COSMOLOGY (DLMA)}}\\

		\cline{4-9}
		&&& \textbf{$\sum_i \lvert m_i \rvert < 0.12$ eV} & \textbf{$\sum_i \lvert m_i \rvert < 0.11$ eV} & \textbf{$\sum_i \lvert m_i \rvert < 0.14$ eV}  & \textbf{$\sum_i \lvert m_i \rvert < 0.12$ eV}&\textbf{$\sum_i \lvert m_i \rvert < 0.11$ eV}&\textbf{$\sum_i \lvert m_i \rvert < 0.14$ eV}\\
	
		\hline\hline
		A1(NO)&$ {\rm NA}$ &$ {\rm NA}$&$ \checkmark $&$ \checkmark $ & $ \checkmark $ &  $ \times $ & $ \times $ &$ \times $ \\ \hline
		
		B1(NO/IO)  &$ \checkmark/\checkmark$&$ \checkmark/\checkmark$&$ \times/\times$&$ \times/\times$& $ \times/\times$&$ \times/\times$& $ \times/\times$&$ \times/\times$ \\ \hline
		
		B2(NO/IO) &$ \checkmark/\checkmark$&$ \checkmark/\checkmark$ &$ \times/\times$&$ \times/\times$ &$ \checkmark/\times$&$\times/\times$& $ \times/\times$&$ \checkmark/\times$\\ \hline
	
		B3(NO/IO) &$ \checkmark/\checkmark$&$ \checkmark/\checkmark$&$\times/\times$&$\times/\times$ & $\times/\times$&$\times/\times$&$\times/\times$&$\times/\times$\\ \hline
		B4(NO/IO) &$ \checkmark$&$ \checkmark$&$\times/\times$&$\times/\times$ &$\times/\times$ &$\times/\times$ &$\times/\times$& $\times/\times$\\ \hline
	
		C1(IO)   &$ \checkmark/\checkmark$&$ \checkmark/\checkmark$&$\times/\times$&$\times/\times$ & $\times/\times$ &$\times/\times$&$\times/\times$&$\times/\times$\\ \hline\hline

	\end{tabular}
	\caption{Allowed and disallowed classes of two-zero texture considering the bounds from NDBD and cosmology. Here NA $\equiv$ not applicable. The $\checkmark$ and $\times$ symbol are used to denote if the class are allowed (disallowed) by current experimental bounds.}\label{t2}
\end{table}
	
	 \begin{table}
	\centering
	\renewcommand{\arraystretch}{1.5}
	\begin{tabular}{|p{1.95cm}|p{1.6cm}|p{1.6cm}|p{1.6cm}|p{1.6cm}|p{1.6cm}|p{1.6cm}|p{1.6cm}|p{1.6cm}|}
		\hline
		\multirow{3}{1.5cm}{\textbf{Class}} &\textbf{NDBD (LMA)}&\textbf{NDBD (DLMA)}& \multicolumn{3}{c|}{\textbf{ COSMOLOGY (LMA)}} & \multicolumn{3}{c|}{\textbf{ COSMOLOGY (DLMA)}}\\
		\cline{4-9}
		&&& \textbf{$\sum_i \lvert m_i \rvert < 0.12$ eV} & \textbf{$\sum_i \lvert m_i \rvert < 0.11$ eV} & \textbf{$\sum_i \lvert m_i \rvert < 0.14$ eV}  & \textbf{$\sum_i \lvert m_i \rvert < 0.12$ eV}&\textbf{$\sum_i \lvert m_i \rvert < 0.11$ eV}&\textbf{$\sum_i \lvert m_i \rvert < 0.14$ eV}\\

		\hline\hline
		G1(NO)&$ {\rm NA}$ &$ {\rm NA}$&$ \checkmark $&$ \checkmark $ & $ \checkmark $ &  $ \times $ & $ \times $ &$ \times $ \\ \hline
	
		G2(NO/IO)  &$ \checkmark/\checkmark$&$ \checkmark/\checkmark$&$ \checkmark/\checkmark$&$ \checkmark/\checkmark$& $ \checkmark/\checkmark$ &$ \checkmark/\checkmark$& $ \checkmark/\checkmark$&$ \checkmark/\checkmark$ \\ \hline
	
		G3(NO/IO) &$ \checkmark/\checkmark$&$ \checkmark/\checkmark$ &$ \checkmark/\checkmark$&$ \checkmark/\checkmark$ &$ \checkmark/\checkmark$&$ \checkmark/\checkmark$& $ \checkmark/\checkmark$&$ \checkmark/\checkmark$\\ \hline

		G4(NO/IO) &$ \checkmark/\checkmark$&$ \checkmark/\checkmark$&$ \times/\checkmark$&$ \times/\checkmark$ & $ \times/\checkmark$&$ \times/\checkmark$& $ \times/\checkmark$&$ \times/\checkmark$\\ \hline
		G5(IO) &$ \checkmark$&$ \checkmark$&$ \checkmark$&$ \checkmark$ & $ \checkmark$ &$ \checkmark$ & $ \checkmark$&  $ \checkmark$\\ \hline

		G6(NO/IO)   &$ \checkmark/\checkmark$&$ \checkmark/\checkmark$&$ \times/\checkmark$&$ \checkmark/\checkmark$ & $ \checkmark/\checkmark$ &$ \times/\checkmark$& $ \checkmark/\checkmark$&$ \checkmark/\checkmark$\\ \hline\hline

	\end{tabular}
\caption{Allowed and disallowed classes of one-zero texture considering the bounds from NDBD and cosmology. Here NA $\equiv$ not applicable. The $\checkmark$ and $\times$ symbol are used to denote if the class are allowed(disallowed) by current experimental bounds.}\label{t3}
\end{table}

As discussed above, constraints from neutrino oscillation experiments, neutrinoless double beta decay experiments and cosmological bound on light neutrino masses allow only one two-zero texture while all the six possible one-zero textures are either partially or fully consistent with all such constraints. 

\section{3+1 $\nu$ Scenario}
\label{sec:level6}
In this section, we check the implications of DLMA solution on Majorana neutrino textures of $3+1$ neutrino scenario. As mentioned earlier, there have been several tantalising hints from experiments like LSND \cite{Aguilar:2001ty} and MiniBooNE \cite{AguilarArevalo:2007it,AguilarArevalo:2010wv,Aguilar-Arevalo:2018gpe} suggesting the presence of additional light neutrinos around eV scale. A few other experiments \cite{Anselmann:1994ar, Abdurashitov:1996dp, Mention:2011rk, Acero:2007su, Giunti:2010zu} have also suggested similar light additional neutrinos. These anomalies received renewed attention recently after the MiniBooNE collaboration reported their new analysis incorporating twice the size data sample than before \cite{Aguilar-Arevalo:2018gpe}, confirming the anomaly at $4.8\sigma$ significance level which becomes $> 6\sigma$ effect if combined with LSND. Previous studies on textures of $3+1$ neutrino scenario have been done in several works \cite{Ghosh:2012pw,Ghosh:2013nya,Zhang:2013mb,Nath:2015emg,Borah:2016xkc,Borah:2017azf, Sarma:2018bgf}.

Evidently, in 3+1 neutrino scenario, the leptonic mixing matrix becomes $4\times 4$. It is well known that $4\times4$ unitary mixing matrix can be parametrised as 

\begin{equation}
U = R_{34} \tilde R_{24} \tilde R_{14}R_{23}\tilde R_{13}R_{12}P
\end{equation}
where 
\begin{equation}
R_{34}= \begin{pmatrix}
1 & 0 & 0 & 0  \\
0 & 1 & 0 & 0  \\
0 & 0 & c_{34} & s_{34}  \\
0 & 0 & -s_{34} & c_{34} \\ 
\end{pmatrix} 
\end{equation}

\begin{equation}
\tilde R_{14}= \begin{pmatrix}
c_{14} & 0 & 0 & s_{14}e^{-i\delta_{14}} \\
0 & 1 & 0 & 0  \\
0 & 0 & 1 & 0  \\
-s_{14}e^{i \delta_{14}} & 0 & 0 & c_{14} \\ 
\end{pmatrix} 
\end{equation}\\
with $c_{ij} = \cos{\theta_{ij}}$, $s_{ij} = \sin{\theta_{ij}}$ , $\delta_{ij}$ being the Dirac CP phases, and 
$$ P = \text{diag}(1, e^{-i\frac{\alpha}{2}}, e^{-i(\frac{\beta}{2}-\delta_{13})}, e^{-i(\frac{\gamma}{2}-\delta_{14})})$$
is the diagonal phase matrix containing the three Majorana phases $\alpha, \beta, \gamma$. In this parametrisation, the six CP phases vary from $-\pi$ to $\pi$. Using the above form of mixing matrix, the $4\times4$ complex symmetric Majorana light neutrino mass matrix can be written as
\begin{eqnarray}
M_{\nu} &=& U M^{\text{diag}}_{\nu} U^T \\
        &=&\begin{pmatrix}
m_{ee} & m_{e\mu} & m_{e\tau} & m_{es}\\
m_{\mu e} & m_{\mu\mu} & m_{\mu\tau} & m_{\mu s} \\
m_{\tau e} & m_{\tau\mu} & m_{\tau\tau} & m_{\tau s} \\
m_{se} & m_{s\mu} & m_{s\tau} & m_{ss}
\end{pmatrix},
\label{mnu}
\end{eqnarray}
 where $M_{\nu}^{diag} = \text{diag}(m_{1}, m_{2}, m_{3}, m_{4})$ 
 is the diagonal light neutrino mass matrix. For normal ordering of active neutrinos i.e., $m_4 > m_3 > m_2 > m_1$, the neutrino mass eigenvalues can be written in terms of the lightest neutrino mass $m_1$ as
  $$m_{2}=\sqrt{m_{1}^{2}+\Delta m_{21}^{2}},\quad  m_{3}=\sqrt{m_{1}^{2}+\Delta m_{31}^{2}},\quad  m_{4}=\sqrt{m_{1}^{2}+\Delta m_{41}^{2}}. $$
Similarly for inverted ordering of active neutrinos i.e., $m_4 > m_2 > m_1 > m_3$, the neutrino mass eigenvalues can be written in terms of the lightest neutrino mass $m_3$ as
 $$m_{1}=\sqrt{m_{3}^{2}-\Delta m_{32}^{2}-\Delta m_{21}^{2}},\quad  m_{2}=\sqrt{m_{3}^{2}-\Delta m_{32}^{2}},\quad  m_{4}=\sqrt{m_{3}^{2}+\Delta m_{43}^{2}}.$$
 Using these, one can analytically write down the $4\times4$ light neutrino mass matrix in terms of three mass squared differences, lightest neutrino mass $m_1 (m_3)$, six mixing angles i.e., $
\theta_{13}$, $\theta_{12}$, $\theta_{23}$, $\theta_{14}$, $\theta_{24}$,
$\theta_{34}$, three Dirac type CP phases i.e., $\delta_{13}$, $\delta_{14}$, 
$\delta_{24}$ and three Majorana type CP phases i.e., $\alpha$, $\beta$, 
$\gamma$.
The analytical expressions of the $4\times4$ light neutrino mass matrix elements are given in Appendix \ref{appen2}. Global fit values of some of the sterile neutrino parameters are given in table \ref{t00} where $\Delta m^2_{\rm LSND} \equiv \Delta m_{41}^{2} (\rm NO), \Delta m_{43}^{2} (\rm IO)$.

Since the light neutrino mass matrix is $4\times 4$, therefore we have many possible texture zeros. As shown in several earlier works \cite{Ghosh:2012pw,Ghosh:2013nya,Zhang:2013mb,Nath:2015emg,Borah:2016xkc,Borah:2017azf}, such texture zeros can not arise in active-sterile or sterile-sterile sector namely $m_{\alpha s} \neq 0, \alpha = e, \mu, \tau, s$. Therefore, only the active $3\times 3$ block of the $4\times 4$ mass matrix can have zeros. Even then, there are many possibilities of one-zero, two-zero, three-zero, four-zero and five-zero as discussed in above mentioned works. Since our purpose is to check the implications of DLMA only, we pick only the most constrained textures namely, four-zero and five-zero textures to check their validity with LMA and DLMA. Although the same has been done for LMA \cite{Borah:2016xkc,Borah:2017azf}, here we check their validity with more updated global fit values of light neutrino parameters.

Accordingly, we have only fifteen possible four-zero textures in 3+1 scenario which are being categorised as class H1 to H10 ($m_{ee}=0$) and H11 to H15 ($m_{ee}\neq 0$) shown in equations \eqref{eq17} to \eqref{eq21}. 
Similarly, we have six phenomenologically allowed five-zero textures shown in equations \eqref{eq22}, \eqref{eq23}. Six-zero texture will have the only one possibility where the entire $3\times3$ active neutrino block of the $4\times 4$ mass matrix will be filled with zeros.

\begin{equation}\label{eq17}
H1=\left(\begin{array}{cccc}
0& 0 &\times&\times\\
0& 0& \times&\times\\
\times & \times& 0&\times\\
\times & \times &\times&\times\\
\end{array}\right),H2=\left(\begin{array}{cccc}
0 & \times&0&\times \\
\times & 0& \times&\times\\
0& \times&0&\times\\
\times & \times &\times&\times\\
\end{array}\right), H3=\left(\begin{array}{cccc}
0 & \times& \times&\times \\
\times &0 & 0&\times\\
\times & 0 & 0&\times\\
\times & \times &\times&\times\\
\end{array}\right)
\end{equation}

\begin{equation}\label{eq18}
H4=\left(\begin{array}{cccc}
0 & 0& 0 &\times\\
0 & 0 & \times&\times\\
0 & \times & \times&\times\\
\times & \times &\times&\times\\
\end{array}\right),H5=\left(\begin{array}{cccc}
0 & 0 &\times&\times\\
0& 0& 0&\times\\
\times & 0& \times&\times\\
\times & \times &\times&\times\\
\end{array}\right),H6=\left(\begin{array}{cccc}
0 & \times&0&\times \\
\times & 0& 0&\times\\
0& 0& \times&\times\\
\times & \times &\times&\times\\
\end{array}\right)
\end{equation}
\begin{equation}\label{eq19}
H7=\left(\begin{array}{cccc}
0& 0& 0&\times \\
0 &\times & \times&\times\\
0 & \times & 0&\times\\
\times & \times &\times&\times\\
\end{array}\right), H8=\left(\begin{array}{cccc}
0& 0& \times &\times\\
0 & \times & 0&\times\\
\times & 0& 0&\times\\
\times & \times &\times&\times\\
\end{array}\right),H9=\left(\begin{array}{cccc}
0 & \times &0&\times\\
\times& \times& 0&\times\\
0 & 0& 0&\times\\
\times & \times& \times&\times\\
\end{array}\right)
\end{equation}
\begin{equation}\label{eq20}
H10=\left(\begin{array}{cccc}
0 & 0&0&\times \\
0 & \times& 0&\times\\
0 & 0& \times&\times\\
\times & \times& \times&\times\\
\end{array}\right), H11=\left(\begin{array}{cccc}
\times & 0& \times&\times \\
0 &0 & 0&\times\\
\times & 0& 0&\times\\
\times & \times& \times&\times\\
\end{array}\right), H12=\left(\begin{array}{cccc}
\times & \times&0&\times\\
\times & 0 & 0&\times\\
0 & 0& 0&\times\\
\times & \times& \times&\times\\
\end{array}\right)
\end{equation}
\begin{equation}\label{eq21}
H13=\left(\begin{array}{cccc}
\times & 0 &0&\times\\
0& 0& \times&\times\\
0 & \times& 0&\times\\
\times & \times& \times&\times\\
\end{array}\right),H14=\left(\begin{array}{cccc}
\times & 0&0&\times \\
0 & 0& 0&\times\\
0 & 0& \times&\times\\
\times & \times& \times&\times\\
\end{array}\right), H15=\left(\begin{array}{cccc}
\times & 0& 0&\times \\
0 &\times & 0&\times\\
0 & 0& 0&\times\\
\times & \times& \times&\times\\
\end{array}\right)
\end{equation}

\begin{equation}\label{eq22}
J1=\left(\begin{array}{cccc}
0& 0 &0&\times\\
0& 0& 0&\times\\
0 & 0& \times &\times\\
\times & \times &\times&\times\\
\end{array}\right),J2=\left(\begin{array}{cccc}
0 & 0&0&\times \\
0 & 0& \times&\times\\
0& \times&0&\times\\
\times & \times &\times&\times\\
\end{array}\right), J3=\left(\begin{array}{cccc}
0 & 0& 0&\times \\
0 &\times & 0&\times\\
0 & 0 & 0&\times\\
\times & \times &\times&\times\\
\end{array}\right)
\end{equation}

\begin{equation}\label{eq23}
J4=\left(\begin{array}{cccc}
0 & 0& \times &\times\\
0 & 0 & 0&\times\\
\times & 0 & 0&\times\\
\times & \times &\times&\times\\
\end{array}\right)
J5=\left(\begin{array}{cccc}
0& \times &0&\times\\
\times& 0& 0&\times\\
0 & 0&0&\times\\
\times & \times &\times&\times\\
\end{array}\right),J6=\left(\begin{array}{cccc}
\times & 0&0&\times \\
0 & 0& 0&\times\\
0& 0&0&\times\\
\times & \times &\times&\times\\
\end{array}\right)
\end{equation}	

		\begin{table}[h!]
		\centering
		\begin{tabular}{||c|  c||}
			\hline
			Parameters & $3 \sigma$ Range  (NO/IO)\\ \hline\hline\hline
			$\Delta m^2_{\rm LSND} [ \rm eV^2]$ & 0.7-2.5/0.7-2.5\\ \hline\hline
			$\sin^2{\theta_{14}}$ & 0.0098-0.0310/0.0098-0.0310 \\ \hline\hline
			$\sin^2{\theta_{24}}$ & 0.0059-0.0262/0.0059-0.0262 	\\ \hline\hline
			$\sin^2{\theta_{34}}$ & 0-0.0396 /0-0.0396\\ \hline\hline
			
		\end{tabular}
		\caption{Global fit 3$\sigma$ values of 3+1 $\nu$ oscillation parameters \cite{Gariazzo:2017fdh,Borah:2017azf, Deepthi:2019ljo}} \label{t00}
	\end{table}
\begin{table}[!htb]
	
	\begin{minipage}{.5\linewidth}
		
		\centering
		\begin{tabular}{||c||c |  c|| }
			\hline
			Class&	DLMA  & LMA  \\ \hline\hline\hline
		H1(NO/IO)&$ \checkmark /\times$	&$ \checkmark /\times$  \\ \hline\hline
		H2(NO/IO)&$ \checkmark /\times $	&$ \checkmark /\times$ \\ \hline\hline
		H3(NO/IO)&$ \times /\checkmark$ 	&$ \times /\checkmark$\\ \hline\hline
		H4(NO/IO)&$ \checkmark /\times$ 	&$ \checkmark/\times$ \\ \hline\hline
		H5(NO/IO)&$ \checkmark /\times$ 	&$ \checkmark /\times$   \\ \hline\hline
		H6(NO/IO)&$ \checkmark /\checkmark$   &$\checkmark /\checkmark$\\ \hline	\hline\hline
		H7(NO/IO)&$ \checkmark /\times$   &$ \checkmark /\times$\\ \hline	\hline
		H8(NO/IO)&$ \times /\times$  &$ \times /\times$\\ \hline	\hline
		H9(NO/IO)&$ \times/\times$   &$ \times /\times$\\ \hline	\hline
		H10(NO/IO)&$ \checkmark /\times$   &$ \checkmark /\times$\\ \hline	\hline

		\end{tabular}
	\end{minipage}%
	\begin{minipage}{.5\linewidth}
		\centering
		
		\begin{tabular}{||c||c |  c|| }
			\hline
			Class&DLMA&	LMA    \\ \hline\hline\hline
		
		H11(NO/IO)&$ \times /\checkmark$&$ \times /\checkmark$ \\ \hline\hline
		H12(NO/IO)&$ \times /\checkmark$&$ \times /\checkmark$	 \\ \hline\hline
		H13(NO/IO)&$ \checkmark /\checkmark$ &$ \checkmark /\checkmark$ \\ \hline\hline
		H14(NO/IO)&$ \times /\checkmark$ &$ \times/\checkmark$  \\ \hline\hline
		H15(NO/IO)&$ \times /\checkmark$ &$ \times /\checkmark$    \\ \hline\hline
			
		\end{tabular}
	\end{minipage} 
	\caption{Summary of allowed and disallowed four zero textures ($m_{ee}=0$) (left) and ($m_{ee}\neq0$) (right) considering LMA and DLMA solution. The $\checkmark$ or $\times$ symbol are used to denote if the class are allowed or disallowed by current experimental bounds.}\label{t6}
\end{table}

	
		
	
\begin{figure}[h]
	
	\includegraphics[width=0.47\textwidth]{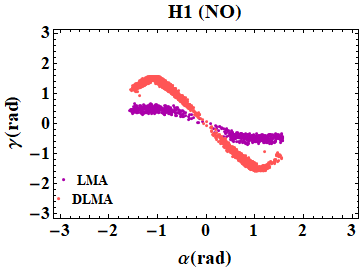}
	\includegraphics[width=0.47\textwidth]{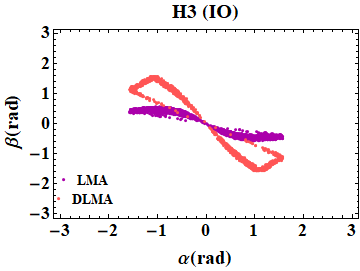}
	\includegraphics[width=0.47\textwidth]{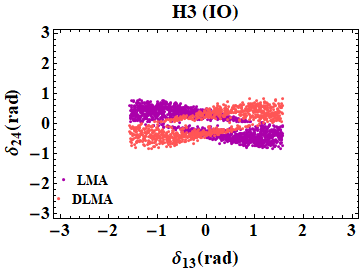}
	\includegraphics[width=0.47\textwidth]{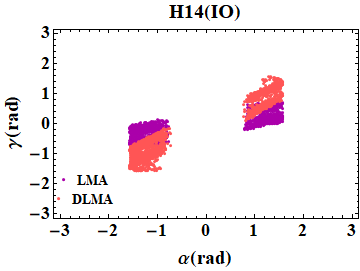}
		\caption{ Correlations between light neutrino parameters for different allowed classes for four-zero texture in $3+1$ neutrino scenario.} \label{fig6}
\end{figure}	
\begin{figure}[h]
	\includegraphics[width=0.47\textwidth]{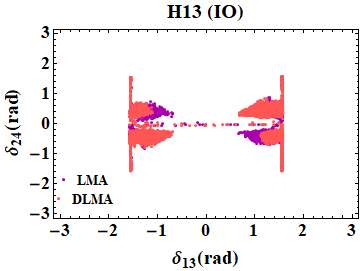}
	\includegraphics[width=0.47\textwidth]{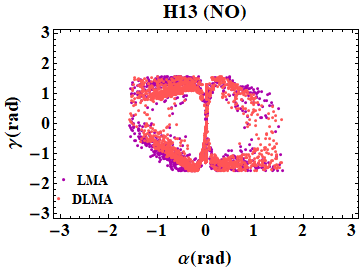}
\includegraphics[width=0.47\textwidth]{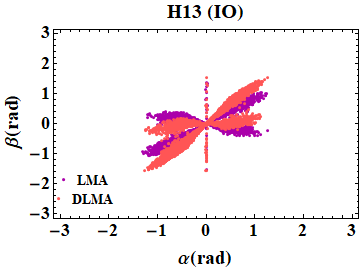}

	\caption{ Correlations between light neutrino parameters for different allowed classes for four-zero texture in $3+1$ neutrino scenario.} \label{fig6a}
\end{figure}

\begin{figure}[h]
	\includegraphics[width=0.47\textwidth]{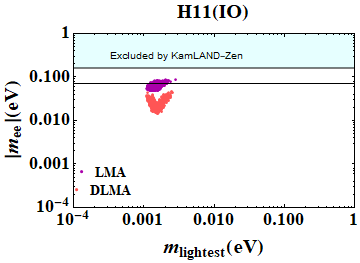}
	\includegraphics[width=0.47\textwidth]{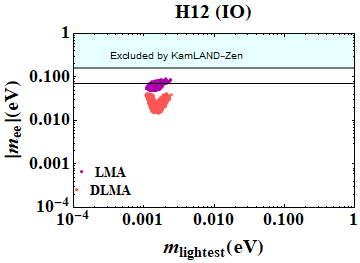}
		\caption{Effective Majorana neutrino mass governing NDBD as a function of the lightest neutrino mass for different allowed classes for four-zero texture.} \label{fig7}
\end{figure}
\begin{figure}[h]
	\includegraphics[width=0.47\textwidth]{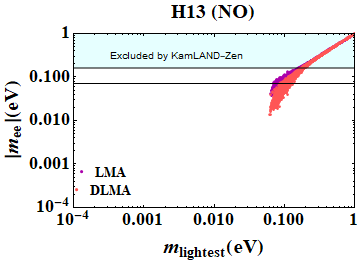}
	\includegraphics[width=0.47\textwidth]{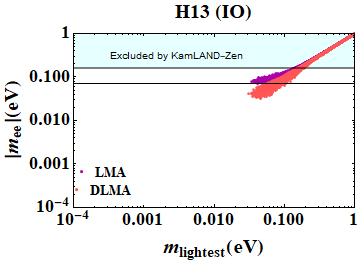}
	\includegraphics[width=0.47\textwidth]{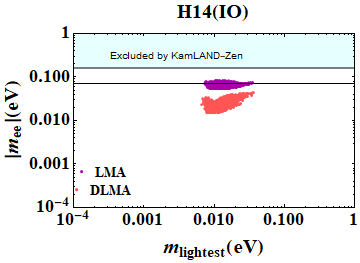}
	\includegraphics[width=0.47\textwidth]{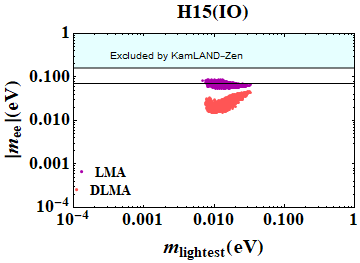}
	\caption{Effective Majorana neutrino mass governing NDBD as a function of the lightest neutrino mass for different allowed classes for four-zero texture.} \label{fig7a}
\end{figure}
We first check the validity of these texture zeros in $3+1$ scheme. Six-zero texture has already been shown to be disallowed while only one of the five-zero textures namely, $J2$ was shown to be allowed in earlier work \cite{Borah:2017azf}. In another earlier work \cite{Borah:2016xkc} where four-zero textures were analysed, it was shown that H3 (NO), H9 (IO), H10 (IO), H11 (NO), H12 (NO) are disallowed. Here we recheck these results in view of the more recent global fit data as well as DLMA solution. Like previous work, we also find the six-zero texture to be disallowed. From the five-zero texture conditions, we have ten real equations, thus we can solve for ten real parameters varying the rest six (five active neutrino parameters and $\Delta m^2_{\rm LSND}$ ) parameters in their 3 $\sigma$ range. We have solved for the six phases, 3 active sterile mixing angles and the lightest neutrino mass. However, interestingly, we find all the cases of five-zero texture for both LMA and DLMA to be disallowed by the latest global fit data.

We use the $3\sigma$ range of sterile neutrino parameters shown in table \ref{t00} while the active neutrino parameters are taken from table \ref{t0} as before. In reference \cite{Gariazzo:2017fdh}, an updated fit of SBL neutrino oscillation data in the 3+1 scenario has been presented wherein the results of the pragmatic 3+1 global fit “PrGlo17”, which includes the MINOS, IceCube and NEOS data is considered as the current best-fit which we have used in our analysis. It should however be noted that more recent studies find the simple $3+1$ neutrino scenario to be in tension with several experiments and additional new physics like sterile neutrino decay have been invoked to find a better fit, as discussed recently by the authors of \cite{Diaz:2019fwt, Moulai:2019gpi}. While MiniBooNE experiment continues to report the excess with more data \cite{Aguilar-Arevalo:2020nvw}, a consistent picture is still missing and future data as well as analysis should shed more light on it. We however, use the available global fit data of $3+1$ neutrino oscillation parameters as mentioned before to see the impact of DLMA solution on possible texture zeros.

Four-zero textures give rise to eight real equations. Out of the sixteen parameters, we can solve for the eight parameters while varying the others in the experimental ranges. For four-zero texture, we solve for six CP phases, lightest neutrino mass and $\theta_{34}$ using the eight real equations while use the $3\sigma$ global fit range of other parameters from table \ref{t0} and \ref{t00}. We summarise our results in table \ref{t6}. As can be seen from this table, LMA or DLMA does not make any distinction as far as allowed and disallowed textures are concerned. Within textures with $m_{ee}=0$ , H8, H9 is completely ruled out while others are allowed only for a particular mass hierarchy. In textures with $m_{ee}\neq 0$, H13 is allowed for both LMA, DLMA as well as NO and IO, the other textures namely H11, H12, H14, H15 are allowed only for IO of light neutrino masses. This agrees only partially with earlier results on four-zero textures \cite{Borah:2016xkc} due to the use of more recent global fit neutrino data. Although, it is not possible to discriminate between LMA and DLMA from the summary of four-zero texture results shown in table \ref{t6}, they give rise to different correlation between neutrino parameters, some of which are shown in figure \ref{fig6} and figure \ref{fig6a}.

In the 3+1 $\nu$ scheme, the effective mass governing NDBD is given by
\begin{equation}
m_{ee}={c^2_{12}} {c^2_{13}} {c^2_{14}} m_1+{c^2_{13}} {c^2_{14}}{s^2_{12}} e^{-i\alpha} m_2+{c^2_{14}} {s^2_{13}} e^{-i\zeta} m_3+{s^2_{14}} e^{-i\gamma} m_4.
\end{equation}
We use the values of these parameters appearing in the expression for $m_{ee}$ as predicted by texture zero conditions and plot the variation of $ \lvert m_{ee} \rvert$ with lightest neutrino mass for allowed four-zero textures of class H11 to H15. The results are shown in figure \ref{fig7} and \ref{fig7a} while the summary is given in table \ref{t7}. None of these textures are ruled out by NDBD bounds although H13 can saturate the bounds for some region of parameter space. On the other hand, H11, H12, H14, H15 can saturate the experimental bound only for LMA. The corresponding predictions for DLMA remains slightly below the current bound but should be within reach of near future experiments. It should be noted that we have not shown the cosmology bounds on neutrino mass in $3+1$ neutrino scenario. Existence of a sterile neutrino with sizeable active-sterile mixing is in conflict with standard cosmology due to the upper bound on sum of absolute neutrino mass mentioned earlier as well as the upper limit on effective relativistic degrees of freedom $N_{\text{eff}} = 2.99 \pm 0.17$ at $68\%$ confidence level (CL) \cite{Aghanim:2018eyx} which is consistent with the SM prediction $N_{\text{eff}} = 3.046$ for three light neutrinos. However, there exists varieties of possibilities of beyond standard model physics (see for example \cite{Dasgupta:2013zpn} where hidden sector interactions of neutrinos are considered) which can alleviate such stringent limits and hence we do not discuss bounds from cosmology here.

	\begin{table}[h]
		
		\begin{tabular}{||c||c |c | c| c|| }
			\hline
			Class&	NDBD(LMA)& NDBD(DLMA)  \\ \hline\hline\hline
			
			H11(IO)&$  \checkmark $ &$ \checkmark $\\ \hline\hline
			H12(IO)&$ \checkmark $&$ \checkmark $\\ \hline\hline
			H13(NO/IO)&$ \checkmark (\checkmark )$&$ \checkmark (\checkmark )$\\ \hline	\hline
			H14(IO)&$ \checkmark $&$ \checkmark $\\ \hline	\hline
			H15(IO)&$ \checkmark $&$ \checkmark $ \\ \hline	\hline
			
		\end{tabular}
		\caption{Allowed classes of four-zero texture considering the bounds from NDBD. The $\checkmark$ symbol is used to denote if the class are allowed by current experimental bounds.}\label{t7}
	\end{table}	
	 \begin{table}
		\centering
		\renewcommand{\arraystretch}{1.3}
		\begin{tabular}{|p{1.5cm}|p{3.5cm}|p{3.5cm}|p{3.5cm}|p{3.5cm}|}
			\hline
			\multirow{2}{1.5cm}{\textbf{Class}} & \multicolumn{2}{c|}{\textbf{LMA}} & \multicolumn{2}{c|}{\textbf{DLMA}}\\
			\cline{2-5}
			& \textbf{$\delta (\rm rad)$} & \textbf{$\sin^2\theta_{23}$} & \textbf{$\delta (\rm rad)$}  & \textbf{$\sin^2\theta_{23}$}\\
			\hline
			G1 (NO) & -1.570-1.570 & 0.427-0.608 &  NA & NA  \\ \hline
			G2 (NO) & (-1.570)-1.570& 0.427-0.609& (-1.570)-1.570 & 0.427-0.609\\ 
			G2 (IO)  &(-1.570)-(-0.44), 0.44-1.570& 0.427-0.608 &(-1.570)-(-0.40), 0.40-1.5707& 0.430-0.612 \\ \hline
			G3 (NO) & (-1.570)-1.570 & 0.427-0.608 & (-1.570)-1.570 & 0.427-0.608 \\ 
			G3 (IO)  &(-1.570)-(-0.451), 0.454-1.570 & 0.427-0.608 &(-1.570)-0.413, 0.408-1.570)& 0.427-0.608\\ \hline
			G4 (IO) &(-1.570)-1.570 & 0.427-0.608 &(-1.570)-1.570 & 0.427-0.608\\ \hline
			G5 (IO) & (-1.570)-1.570 & 0.427-0.608 &(-1.570)-1.570 & 0.427-0.608  \\ \hline
			
			G6 (IO)   &(-1.570)-1.570 & 0.427-0.608 &(-1.570)-1.570& 0.427-0.608\\ \hline\hline\hline
			A1 & (-1.570) -1.570 & 0.427-0.608& NA  &  NA \\ \hline
%
%
			 
		\end{tabular}
		\caption{Predicted range of atmospheric mixing angle and Dirac CP phase in allowed one-zero and two-zero textures of three neutrino scenario. Here NA $\equiv$ not allowed. }\label{t8}
	\end{table}

	\begin{table}
	\centering
	\renewcommand{\arraystretch}{1.3}
	\begin{tabular}{|p{1.7cm}|p{3.55cm}|p{3.5cm}|p{3.55cm}|p{3.5cm}|}
		\hline
		\multirow{2}{1.5cm}{\textbf{Class}} & \multicolumn{2}{c|}{\textbf{LMA}} & \multicolumn{2}{c|}{\textbf{DLMA}}\\
		\cline{2-5}
		& \textbf{$\delta_{13} (\rm rad)$ } & \textbf{$\sin^2\theta_{23}$} & \textbf{$\delta_{13} (\rm rad)$ }  & \textbf{$\sin^2\theta_{23}$}\\
		\hline
		H1 (NO) & (-1.568)-1.569 & 0.427-0.608 &(-1.568)-1.562  &  0.427-0.608\\ \hline
		H2 (NO) & (-1.568)-1.569& 0.427-0.608& -1.548-1.568 & 0.427-0.608 \\ 
		\hline
		 
 	  H3 (IO) &(-1.5705)-1.562& 0.430-0.611 &(-1.570)-1.569& 0.430-0.611 \\ \hline
		H4 (NO) & (-1.570)-1.567 & 0.427-0.608& (-1.568)-1.568 & 0.427-0.608\\ \hline
	 
		H5 (NO) & (-1.570)-1.55 & 0.516-0.608 &(-1.515)-1.547 & 0.451-0.608  \\ \hline
		
		H6 (NO) & (-1.337)-1.429 & 0.494-0.602 & (-1.402)-1.407 &0.495-0.608  \\ 
		H6 (IO) &(-0.641)-1.347& 0.550-0.610 &(-1.483)-1.373& 0.523-0.606 \\ \hline
		H7 (NO) & (-1.564)-(-1.516) & 0.428-0.607&0.505-0.506  &0.600-0.602  \\ \hline
			
		H10 (NO) & (-1.43)-1.408 & 0.477-0.598 & (-1.5707)-1.703 & 0.427-0.608  \\ \hline

		H11 (IO)&(-1.56)-1.55 &0.430-0.611 &( -1.545)-1.56&0.43-0.61 \\ \hline

	H12 (IO)&(-1.569)-1.565&0.430-0.611&(-1.567)-1.570 &0.430-0.611 \\ \hline
	H13 (NO) &  (-1.570)-1.570 & 0.427-0.608&(-1.569)-1.570&0.427-0.608  \\ 
	H13 (IO)&(-1.570)-1.570 & 0.427-0.608 &(-1.570)-1.570& 0.427-0.608 \\ \hline

	H14 (IO)&(-1.566)-1.569&0.430-0.611&(-1.569)-1.566&0.430-0.611 \\ \hline

	H15 (IO)&(-1.567)-1.570&0.430-0.611&(-1.569)-1.555& 0.430-0.611 \\ \hline

	\end{tabular}
	\caption{Predicted range of atmospheric mixing angle and one of the Dirac CP phases in allowed four-zero textures of $3+1$ neutrino scenario.}\label{t8a}
\end{table}

\section{Conclusion}{\label{sec:level7}}
We have studied studied the possibility of texture zeros in Majorana light neutrino mass matrix in the light of Dark LMA solution to solar neutrino problem where solar mixing angle $\sin^2{\theta_{12}}\simeq 0.7 $ lies in the second octant. In order to make a comparison with the standard LMA solution, we check the validity of different possible texture zero scenarios namely one-zero and two-zero textures in three neutrino scenarios using both LMA and DLMA solutions. We find that using the latest global fit data for three neutrino scenario and cosmological upper bound on sum of absolute neutrino masses, all two-zero textures with DLMA are ruled out, except for B2 (NO) which satisfies the cosmological bound on sum of absolute neutrino mass, $\sum_i \lvert m_i \rvert <  0.14 \; {\rm eV}$ . With LMA however, one possible two-zero texture (out of fifteen possibilities) denoted by A1 is still allowed. One the other hand, one-zero textures are less restricted compared to two-zero textures. Using all available constraints from neutrino data, neutrinoless double beta decay and cosmology, five out of six possible one-zero textures are allowed only with IO of light neutrino mass. G1 is allowed only with LMA and NO while three (G2, G3, G6) are allowed for both the hierarchies as well as LMA, DLMA.

Apart from such differences between LMA, DLMA  as well as between mass hierarchies leading to allowed and disallowed texture zeros, we also get interesting correlations between light neutrino parameters for allowed cases which distinguish LMA from DLMA. Such correlations or specific predictions of light neutrino parameters like Dirac CP phase, octant of atmospheric angle, neutrino mass ordering can be probed at ongoing as well as upcoming neutrino oscillation experiments. We summarize these predictions for the allowed one-zero and two-zero textures in table \ref{t8}. While there is no preference shown for particular octant of atmospheric mixing angle, the textures G2 (IO) and G3 (IO) are inconsistent with vanishing Dirac CP phases, which is also suggested by recent neutrino oscillation experiments \cite{Abe:2019vii}. Also most of these textures also saturate the experimental limit on neutrinoless double beta decay amplitude, keeping them within reach of upcoming experiments.

Finally we extend our studies in 3 neutrino scenario to $3+1$ neutrino scenario by focusing on the most constrained scenarios namely six-zero, five-zero and four-zero scenarios. While we find the six-zero and all five-zero textures to be disallowed in view of recent global fit data with both LMA and DLMA, a few of the four-zero textures are found to be allowed from neutrino oscillation data as well as neutrinoless double beta decay constraints. While LMA or DLMA does not play a decisive role in $3+1$ neutrino case (unlike in 3 neutrino scenario), they do give rise to different predictions for light neutrino parameters, apparent from their correlation plots. We summarise the predictions for atmospheric mixing angle and one of the Dirac CP phases in table \ref{t8a}. These textures and predictions of light neutrino parameters can be tested at neutrino oscillation and neutrinoless double beta decay experiments. 

To summarise, our study not only compares LMA and DLMA solutions to solar neutrino problem from Majorana neutrino textures point of view in 3 neutrino as well as 3+1 neutrino scenarios but also gives an update on the validity of these textures with the standard LMA solution. Many of these textures found to be allowed in both these scenarios by earlier studies have now been found to be disallowed due to stringent constraints from neutrino oscillation data, neutrinoless double beta decay as well as cosmology. More stringent data from future experiments should be able to reduce the number of such possibilities further.
	
\acknowledgments
The authors acknowledge the support from Early Career Research Award from the department of science and technology-science and engineering research board (DST-SERB), Government of India (reference number: ECR/2017/001873) 

\appendix

\section{Light neutrino mass matrix elements in 3 neutrino scenario}
\label{appen1}
{\small \begin{widetext}
\begin{equation}
M_{ee} = c^2_{12}c^2_{13} m_1+c^2_{13} s^2_{12} m_2 e^{i 2\alpha} + s^2_{13} m_3 e^{i2\beta}
\end{equation}
\begin{align}
M_{e \mu}=M_{\mu e} &= c_{13}\bigg(s_{13}s_{23}m_3 e^{i(\delta_{\text{cp}}+2\beta)}-c_{12}m_1(c_{23}s_{12}+c_{12}s_{13}s_{23}e^{i\delta_{\text{cp}}}) \nonumber \\
& +s_{12}m_2 e^{i2\alpha} (c_{12}c_{23}-s_{12}s_{13}s_{23}e^{i\delta_{\text{cp}}}) \bigg) 
\end{align}
\begin{align}
M_{e\tau}=M_{\tau e}  &= c_{13} \bigg ( c_{23}s_{13}m_3 e^{i(\delta_{\text{cp}}+2\beta)}-s_{12}m_2 e^{i2\alpha} (c_{23}s_{12}s_{13}e^{i\delta_{\text{cp}}} \nonumber \\
& +c_{12}s_{23})+c_{12}m_1(-c_{12}c_{23}s_{13}e^{i\delta_{\text{cp}}}+s_{12}s_{23})\bigg )
\end{align}
\begin{equation}
M_{\mu \mu} = c^2_{13} s^2_{23} m_3 e^{i2(\delta_{\text{cp}}+\beta)}+m_1 (c_{23}s_{12} +c_{12} s_{13} s_{23} e^{i\delta_{\text{cp}}})^2+m_2 e^{i2\alpha}(c_{12}c_{23}-s_{12}s_{13}s_{23}e^{i\delta_{\text{cp}}})^2
\end{equation}
\begin{align}
M_{\mu \tau} =M_{\tau \mu} &= c^2_{13}c_{23}s_{23}m_3 e^{i2(\delta_{\text{cp}}+\beta)}+m_1 (c_{12}c_{23}s_{13} e^{i\delta_{\text{cp}}}-s_{12}s_{23})(c_{23}s_{12}+c_{12}s_{13}s_{23}e^{i\delta_{\text{cp}}}) \nonumber \\
&-m_2e^{i2\alpha}(c_{23}s_{12}s_{13}e^{i\delta_{\text{cp}}}+c_{12}s_{23})(c_{12}c_{23}-s_{12}s_{13}s_{23}e^{i\delta_{\text{cp}}})
\end{align}
\begin{equation}
M_{\tau \tau} = c^2_{13}c^2_{23}m_3 e^{i2(\delta_{\text{cp}}+\beta)}+m_2e^{i 2\alpha} (c_{23}s_{12}s_{13}e^{i\delta_{\text{cp}}}+c_{12}s_{23})^2+m_1(c_{12}c_{23}s_{13}e^{i\delta_{\text{cp}}}-s_{12}s_{23})^2
\end{equation}

\section{Light neutrino mass matrix elements in 3+1 neutrino scenario}
\label{appen2}

{\small \begin{widetext}
\begin{equation}
M_{ee} = c_{12}^2 c_{13}^2 c_{14}^2 m_1+e^{- i \alpha } c_{13}^2 c_{14}^2 m_2 s_{12}^2+e^{- i \beta } c_{14}^2 m_3 s_{13}^2+e^{-i \gamma } m_4 s_{14}^2 \nonumber
\end{equation}
\begin{eqnarray}
M_{e\mu} &=&-e^{-i \delta _{24}} c_{14} \big(e^{i \delta _{24}} c_{12} c_{13} c_{23} c_{24} \big(m_1-e^{- i \alpha } m_2\big) s_{12}-e^{i \big(\delta
_{13}+\delta _{24}\big)} c_{13} c_{24} \big(e^{- i \beta } m_3-e^{- i \alpha } m_2 s_{12}^2\big) s_{13} s_{23} \nonumber \\
 && +e^{i \big(2 \alpha +\delta _{14}\big)}M
c_{13}^2 m_2 s_{12}^2 s_{14} s_{24}-e^{i \delta _{14}} \big(e^{- i \gamma } m_4-e^{- i \beta } m_3 s_{13}^2\big) s_{14} s_{24}+c_{12}^2 c_{13}
m_1 \big(e^{i \big(\delta _{13}+\delta _{24}\big)} c_{24} s_{13} s_{23} \nonumber  \\
&& +e^{i \delta _{14}} c_{13} s_{14} s_{24}\big)\big) \nonumber
\end{eqnarray}
\begin{eqnarray}
M_{e\tau}&=&c_{14} \big(-e^{i \big(- \alpha +\delta _{14}\big)} c_{13}^2 c_{24} m_2 s_{12}^2 s_{14} s_{34}+e^{i \delta _{14}} c_{24} \big(e^{- i \gamma
} m_4-e^{- i \beta } m_3 s_{13}^2\big) s_{14} s_{34} \nonumber \\
&& +c_{12} c_{13} \big(m_1-e^{- i \alpha } m_2\big) s_{12} \big(c_{34} s_{23}+e^{i \delta
_{24}} c_{23} s_{24} s_{34}\big)+e^{i \delta _{13}} c_{13} \big(e^{- i \beta } m_3-e^{- i \alpha } m_2 s_{12}^2\big) s_{13} \big(c_{23} c_{34} \nonumber \\
&& -e^{i
\delta _{24}} s_{23} s_{24} s_{34}\big)-c_{12}^2 c_{13} m_1 \big(e^{i \delta _{13}} c_{23} c_{34} s_{13}+\big(e^{i \delta _{14}} c_{13} c_{24}
s_{14}-e^{i \big(\delta _{13} +\delta _{24}\big)} s_{13} s_{23} s_{24}\big) s_{34}\big)\big) \nonumber
\end{eqnarray}
\begin{eqnarray}
M_{\mu\mu} &=&e^{ i \big(-\gamma + 2\delta_{14}- 2 \delta_{24}\big)} c_{14}^2 m_4 s_{24}^2+e^{- i \beta } m_3 \big(e^{i \delta _{13}} c_{13} c_{24} s_{23}-e^{i
\big(\delta _{14}-\delta _{24}\big)} s_{13} s_{14} s_{24}\big){}^2+e^{- i \alpha } m_2 \big(c_{12} c_{23} c_{24}\nonumber \\
&& +s_{12} \big(-e^{i \delta
_{13}} c_{24} s_{13} s_{23}-e^{i \big(\delta _{14}-\delta _{24}\big)} c_{13} s_{14} s_{24}\big)\big){}^2+m_1 \big(c_{23} c_{24} s_{12}+c_{12}
\big(e^{i \delta _{13}} c_{24} s_{13} s_{23} \nonumber \\
&& +e^{i \big(\delta _{14}-\delta _{24}\big)} c_{13} s_{14} s_{24}\big)\big){}^2 \nonumber
\end{eqnarray}
\begin{eqnarray}
M_{\mu\tau} &=& e^{i \big(- \gamma +2 \delta _{14}-\delta _{24}\big)} c_{14}^2 c_{24} m_4 s_{24} s_{34}+e^{i \big(2 \beta +\delta _{13}\big)} m_3 \big(e^{i
\delta _{13}} c_{13} c_{24} s_{23}-e^{i \big(\delta _{14}-\delta _{24}\big)} s_{13} s_{14} s_{24}\big)  \nonumber \\
&&\big(-e^{-i \big(\delta _{13}-\delta
_{14}\big)} c_{24} s_{13} s_{14} s_{34}+c_{13} \big(c_{23} c_{34}-e^{i \delta _{24}} s_{23} s_{24} s_{34}\big)\big)+m_1 \big(-c_{23} c_{24}
s_{12}+c_{12} \big(-e^{i \delta _{13}} c_{24} s_{13} s_{23} \nonumber \\
&& -e^{i \big(\delta _{14}-\delta _{24}\big)} c_{13} s_{14} s_{24}\big)\big) \big(s_{12}
\big(c_{34} s_{23}+e^{i \delta _{24}} c_{23} s_{24} s_{34}\big)+c_{12} \big(-e^{i \delta _{14}} c_{13} c_{24} s_{14} s_{34}-e^{i \delta _{13}}
s_{13} \big(c_{23} c_{34} \nonumber \\
&&-e^{i \delta _{24}} s_{23} s_{24} s_{34}\big)\big)\big)+e^{- i \alpha } m_2 \big(c_{12} c_{23} c_{24}+s_{12} \big(-e^{i
\delta _{13}} c_{24} s_{13} s_{23}-e^{i \big(\delta _{14}-\delta _{24}\big)} c_{13} s_{14} s_{24}\big)\big) \big(-c_{12} \big(c_{34} s_{23} \nonumber \\
&&+e^{i
\delta _{24}} c_{23} s_{24} s_{34}\big)+s_{12} \big(-e^{i \delta _{14}} c_{13} c_{24} s_{14} s_{34}-e^{i \delta _{13}} s_{13} \big(c_{23} c_{34}-e^{i
\delta _{24}} s_{23} s_{24} s_{34}\big)\big)\big) \nonumber
\end{eqnarray}
\begin{eqnarray}
M_{\tau \tau} &=& e^{ i \big(-\gamma + 2 \delta_{14}\big)} c_{14}^2 c_{24}^2 m_4 s_{34}^2+e^{ i \big(-\beta +2\delta_{13}\big)} m_3 \big(e^{-i \big(\delta
_{13}-\delta _{14}\big)} c_{24} s_{13} s_{14} s_{34}+c_{13} \big(-c_{23} c_{34}+e^{i \delta _{24}} s_{23} s_{24} s_{34}\big)\big){}^2 \nonumber \\
&& +m_1\big(s_{12} \big(c_{34} s_{23}+e^{i \delta _{24}} c_{23} s_{24} s_{34}\big)+c_{12} \big(-e^{i \delta _{14}} c_{13} c_{24} s_{14} s_{34}-e^{i
\delta _{13}} s_{13} \big(c_{23} c_{34}-e^{i \delta _{24}} s_{23} s_{24} s_{34}\big)\big)\big){}^2 \nonumber \\
&& +e^{- i \alpha } m_2 \big(c_{12} \big(c_{34}
s_{23}+e^{i \delta _{24}} c_{23} s_{24} s_{34}\big)-s_{12} \big(-e^{i \delta _{14}} c_{13} c_{24} s_{14} s_{34}-e^{i \delta _{13}} s_{13} \big(c_{23}
c_{34} -e^{i \delta _{24}} s_{23} s_{24} s_{34}\big)\big)\big){}^2 \nonumber
\end{eqnarray}
\begin{eqnarray}
M_{es} &=& c_{14} \big(e^{i \delta _{14}} c_{24} c_{34} \big(e^{- i \gamma } m_4-e^{- i \alpha } c_{13}^2 m_2 s_{12}^2-e^{- i \beta } m_3 s_{13}^2\big)
s_{14}-e^{i \delta _{13}} c_{13} \big(e^{- i \beta } m_3-e^{- i \alpha } m_2 s_{12}^2\big) s_{13}  \nonumber \\
&& \big(e^{i \delta _{24}} c_{34} s_{23} s_{24}+c_{23}
s_{34}\big)+c_{12} c_{13} \big(m_1-e^{- i \alpha } m_2\big) s_{12} \big(e^{i \delta _{24}} c_{23} c_{34} s_{24}-s_{23} s_{34}\big) \nonumber \\
&& -c_{12}^2
c_{13} m_1 \big(e^{i \delta _{14}} c_{13} c_{24} c_{34} s_{14}-e^{i \delta _{13}} s_{13} \big(e^{i \delta _{24}} c_{34} s_{23} s_{24}+c_{23} s_{34}\big)\big)\big) \nonumber
\end{eqnarray}
\begin{eqnarray}
M_{\mu s} &=& e^{i \big(2 \gamma +2 \delta _{14}-\delta _{24}\big)} c_{14}^2 c_{24} c_{34} m_4 s_{24}+e^{i \big(2 \beta +\delta _{13}\big)} m_3 \big(e^{i
\delta _{13}} c_{13} c_{24} s_{23}-e^{i \big(\delta _{14}-\delta _{24}\big)} s_{13} s_{14} s_{24}\big) \nonumber \\
&& \big(-e^{-i \big(\delta _{13}-\delta
_{14}\big)} c_{24} c_{34} s_{13} s_{14}-c_{13} \big(e^{i \delta _{24}} c_{34} s_{23} s_{24}+c_{23} s_{34}\big)\big)+m_1 \big(-c_{23} c_{24}
s_{12}+c_{12} \big(-e^{i \delta _{13}} c_{24} s_{13} s_{23} \nonumber \\
&&  -e^{i \big(\delta _{14}-\delta _{24}\big)} c_{13} s_{14} s_{24}\big)\big)\big(s_{12}
\big(e^{i \delta _{24}} c_{23} c_{34} s_{24}-s_{23} s_{34}\big)+c_{12} \big(-e^{i \delta _{14}} c_{13} c_{24} c_{34} s_{14}+e^{i \delta _{13}}
s_{13} \big(e^{i \delta _{24}} c_{34} s_{23} s_{24} \nonumber \\
&& +c_{23} s_{34}\big)\big)\big)+e^{- i \alpha } m_2 \big(c_{12} c_{23} c_{24}+s_{12}\big(-e^{i
\delta _{13}} c_{24} s_{13} s_{23}-e^{i \big(\delta _{14}-\delta _{24}\big)} c_{13} s_{14} s_{24}\big)\big) \big(c_{12} \big(-e^{i \delta
_{24}} c_{23} c_{34} s_{24} \nonumber \\
&&+s_{23} s_{34}\big)+s_{12} \big(-e^{i \delta _{14}} c_{13} c_{24} c_{34} s_{14}+e^{i \delta _{13}} s_{13} \big(e^{i
\delta _{24}} c_{34} s_{23} s_{24}+c_{23} s_{34}\big)\big)\big) \nonumber
\end{eqnarray}
\begin{eqnarray}
M_{\tau s} &=& e^{ i \big(-\gamma +2\delta_{14}\big)} c_{14}^2 c_{24}^2 c_{34} m_4 s_{34}+e^{ i \big(-\beta +2 \delta_{13}\big)} m_3 \big(-e^{-i \big(\delta
_{13}-\delta _{14}\big)} c_{24} c_{34} s_{13} s_{14}-c_{13} \big(e^{i \delta _{24}} c_{34} s_{23} s_{24}+c_{23} s_{34}\big)\big)  \nonumber \\
&& \big(-e^{-i
\big(\delta _{13}-\delta _{14}\big)} c_{24} s_{13} s_{14} s_{34}+c_{13} \big(c_{23} c_{34}-e^{i \delta _{24}} s_{23} s_{24} s_{34}\big)\big)+m_1
\big(s_{12} \big(e^{i \delta _{24}} c_{23} c_{34} s_{24}-s_{23} s_{34}\big) \nonumber \\
&& +c_{12} \big(-e^{i \delta _{14}} c_{13} c_{24} c_{34} s_{14}+e^{i
\delta _{13}} s_{13} \big(e^{i \delta _{24}} c_{34} s_{23} s_{24}+c_{23} s_{34}\big)\big)\big) \big(s_{12} \big(c_{34} s_{23}+e^{i \delta
_{24}} c_{23} s_{24} s_{34}\big) \nonumber \\
&& +c_{12} \big(-e^{i \delta _{14}} c_{13} c_{24} s_{14} s_{34}-e^{i \delta _{13}} s_{13} \big(c_{23} c_{34}-e^{i
\delta _{24}} s_{23} s_{24} s_{34}\big)\big)\big)+e^{- i \alpha } m_2 \big(c_{12} \big(-e^{i \delta _{24}} c_{23} c_{34} s_{24}+s_{23} s_{34}\big) \nonumber \\
&& +s_{12}
\big(-e^{i \delta _{14}} c_{13} c_{24} c_{34} s_{14}+e^{i \delta _{13}} s_{13} \big(e^{i \delta _{24}} c_{34} s_{23} s_{24}+c_{23} s_{34}\big)\big)\big)
\big(-c_{12} \big(c_{34} s_{23}+e^{i \delta _{24}} c_{23} s_{24} s_{34}\big) \nonumber \\
&& +s_{12} \big(-e^{i \delta _{14}} c_{13} c_{24} s_{14} s_{34} -e^{i
\delta _{13}} s_{13} \big(c_{23} c_{34}-e^{i \delta _{24}} s_{23} s_{24} s_{34}\big)\big)\big) \nonumber
\end{eqnarray}
\begin{eqnarray}
M_{ss} &=& e^{- i \big(\gamma +\delta _{14}\big)} c_{14}^2 c_{24}^2 c_{34}^2 m_4+e^{ i \big(-\beta + 2 \delta_{13}\big)} m_3 \big(e^{-i \big(\delta
_{13}-\delta _{14}\big)} c_{24} c_{34} s_{13} s_{14}+c_{13} \big(e^{i \delta _{24}} c_{34} s_{23} s_{24}+c_{23} s_{34}\big)\big){}^2 \nonumber \\
&& +m_1 \big(s_{12}
\big(e^{i \delta _{24}} c_{23} c_{34} s_{24}-s_{23} s_{34}\big)+c_{12} \big(-e^{i \delta _{14}} c_{13} c_{24} c_{34} s_{14}+e^{i \delta _{13}}
s_{13} \big(e^{i \delta _{24}} c_{34} s_{23} s_{24} +c_{23} s_{34}\big)\big)\big){}^2 \nonumber \\
&&+e^{- i \alpha } m_2 \big(c_{12} \big(-e^{i \delta _{24}}
c_{23} c_{34} s_{24}+s_{23} s_{34}\big)  \nonumber \\
&& +s_{12} \big(-e^{i \delta _{14}} c_{13} c_{24} c_{34} s_{14}+e^{i \delta _{13}} s_{13} \big(e^{i \delta
_{24}} c_{34} s_{23} s_{24}+c_{23} s_{34}\big)\big)\big){}^2 \nonumber
\end{eqnarray}

\end{widetext}}

\end{widetext}}

\bibliographystyle{JHEP}
\bibliography{TODLMA}

\end{document}